\documentclass[useAMS,usenatbib]{mn2e}

\usepackage{graphicx}
\usepackage{aas_macros}
\usepackage{amsfonts,amssymb}
\usepackage{color}
\usepackage{numdef}

\newcommand{\lya}{Ly$\alpha$}
\newcommand{\lyb}{Ly$\beta$}
\newcommand{\lyg}{Ly$\gamma$}
\newcommand{\hi}{H~{\sc i}}
\newcommand{\hii}{H~{\sc ii}}
\newcommand{\oi}{O~{\sc i}}
\newcommand{\cii}{C~{\sc ii}}

\newcommand{\civ}{C~{\sc iv}}
\newcommand{\siii}{Si~{\sc ii}}

\newcommand{\siiv}{Si~{\sc iv}}
\newcommand{\nv}{N~{\sc v}}

\newcommand{\mgii}{Mg~{\sc ii}}
\newcommand{\feii}{Fe~{\sc ii}}
\newcommand{\alii}{Al~{\sc ii}}
\newcommand{\ovi}{O~{\sc vi}}

\newcommand{\kms}{km~s$^{-1}$}

\newcommand{\taueff}{$\tau_{\rm eff}$}

\newcommand{\taueffa}{$\tau_{\rm eff}^{\alpha}$}
\newcommand{\taueffb}{$\tau_{\rm eff}^{\beta}$}

\newcommand{\hinvMpc}{Mpc$\,h^{-1}$}
\num\newcommand{\ulas0148}{ULAS~J0148$+$0600}
\newcommand{\Ptaueff}{$P($$\le$$\tau_{\rm eff})$}
\newcommand{\fhi}{$\langle f_{\rm H\,I} \rangle$}
\newcommand{\G}{$\Gamma$}

\newcommand{\mfp}{$\lambda_{\rm mfp}^{912}$}

\newcommand{\fion}{$f_{\rm ion}$}

\title[Evidence of patchy reionization]{Evidence of patchy hydrogen reionization from an extreme \lya\ trough below redshift six}

\author[Becker et al.]
   {George D.~Becker$^1$\thanks{gdb@ast.cam.ac.uk}, James S.~Bolton$^2$, Piero Madau$^3$, Max Pettini$^1$, \newauthor Emma V.~Ryan-Weber$^4$ and Bram P.~Venemans$^5$ \\
   $^1$Kavli Institute for Cosmology and Institute of Astronomy, University of Cambridge, Madingley Road, Cambridge, CB3 0HA, UK \\
   $^2$School of Physics and Astronomy, University of Nottingham, University Park, Nottingham NG7 2RD, UK \\
   $^3$Department of Astronomy and Astrophysics, University of California, 1156 High Street, Santa Cruz, CA 95064, USA \\
   $^4$Centre for Astrophysics and Supercomputing, Swinburne University of Technology, Hawthorn, VIC 3122, Australia \\
   $^5$Max-Planck Institute for Astronomy, Kšnigstuhl 17, D-69117 Heidelberg, Germany}
   \date{Draft version \today}   
   
\begin{document}

\label{firstpage}

\maketitle

\begin{abstract}  

We report the discovery of an extremely long ($\sim$110~\hinvMpc)
  and dark ($\tau_{\rm eff} \gtrsim 7$) \lya\ trough extending down to
  $z \simeq 5.5$ towards the $z_{\rm em} \simeq 6.0$ quasar \ulas0148.
  We use these new data in combination with \lya\ forest measurements
  from 42 quasars at $4.5 \le z_{\rm em} \le 6.4$ to conduct an updated analysis of the line-of-sight
  variance in the intergalactic \lya\ opacity over $4 \le z \le 6$.
We find that the scatter in transmission among lines of
sight near $z \sim 6$ significantly exceeds theoretical expectations
for either a uniform ultraviolet background (UVB) or simple
fluctuating UVB models in which the mean free path to ionizing
photons is spatially invariant.  The data, particularly near $z \simeq
5.6$--5.8, instead require fluctuations in the volume-weighted hydrogen
neutral fraction that are a factor of 3 or more beyond
  those expected from density variations alone.  We argue that these fluctuations are
most likely driven by large-scale variations in the mean free path, consistent with expectations for the final stages of  inhomogeneous hydrogen reionization.  Even by $z \simeq
5.6$, however, a large fraction of the data are consistent with a
uniform UVB, and by $z \sim 5$ the data are fully consistent with
opacity fluctuations arising solely from the density field.  This suggests that while reionization may be ongoing at $z \sim 6$, it has fully completed by $z \sim 5$.

\end{abstract}


\section{Introduction}

Determining how and when the intergalactic medium (IGM) became
reionized is currently one of the key goals of extragalactic
astronomy.  Within roughly one billion years of the big bang,
ultraviolet photons from the first luminous objects ionized nearly
every atom in the IGM.  The details of this process reflect the nature
of the first stars, galaxies, and active galactic nuclei (AGN), as well as
the characteristics of large-scale structure, and therefore continue to be the
subject of considerable observational and theoretical effort.

Some of the most fundamental constraints on when reionization ended
come from the evolution of intergalactic \lya\ opacity near $z \sim
6$, as measured in the spectra of high-redshift quasars
\citep[e.g.,][]{becker2001,djorgovski2001,fan2002,fan2006b,white2003,songaila2004}
and gamma-ray bursts \citep[e.g.,][]{chornock2013,chornock2014}.  The
largest data set to date was provided by \cite{fan2006b}, who measured
the opacity in the \lya\ forest towards a sample of 19 $z \sim 6$
quasars.  The fact that transmitted flux is observed in the
\lya\ forest up to $z \sim 6$ suggests that reionization had largely
ended by that point, at least in a volume-averaged sense.  Fan et
al. noted a rapid increase in the mean \lya\ opacity at $z > 5.7$,
however, which suggests a decline in the intensity of the ultraviolet
background (UVB) near 1~Ryd \citep[see
  also][]{bolton2007,calverley2011,wyithe2011}.  They also noted a
large sightline-to-sightline scatter \citep[see also][]{songaila2004},
which they interpreted as evidence of large (factor of $\gtrsim$4)
fluctuations in the UVB near $z \sim 6$.  Further evidence for a decline in the UVB from $z \sim 5$ to 6 is also potentially seen in the changing ionization state of metal-enriched absorbers over this interval \citep{becker2009,rw2009,becker2011b,simcoe2011a,dodorico2013,keating2014}.

The inferred rapid evolution in the UVB over $5 \lesssim z \lesssim 6$
stands in stark contrast to its nearly constant value over $2 < z < 5$
\citep[e.g.,][]{bolton2005,fg2008b,becker2013b}.  It is unclear,
however, whether a rapidly evolving UVB necessarily indicates a recent
end to reionization.  As pointed out by \citet{mcquinn2011}, a modest
increase in the global ionizing emissivity may produce a large
increase in the mean free path to ionizing photons, leading to a
strong increase in the UVB.  Such an evolution may be driven by the
increase in the star-formation rate density from $z \sim 6$ to 5
\citep[e.g.,][]{bouwens2007} even if reionization ended significantly
earlier.  

On the other hand, \cite{lidz2007} and subsequently
  \cite{mesinger2010} have pointed out that existing measurements
  of the intergalactic \lya\ opacity do not firmly
  rule out the final stages of reionization occurring at $z \lesssim 6$.  The spatially inhomogeneous nature of reionization and the limited number of quasar sight-lines available at $z>5$ may
  conspire together such that isolated, neutral patches in the IGM
  remain as yet undetected at $z\sim 5$--6.

In this context, the case for recent (or ongoing) reionization at $z \sim 6$ would be significantly clarified by determining whether the observed scatter in \lya\ opacity at $z \sim 6$ is truly driven by fluctuations in the UVB, as proposed by \citet{fan2006b}.  The claim of large UVB fluctuations was
queried by \citet{lidz2006}, who argued that significant
sightline-to-sightline variations in opacity are expected due to
large-scale density fluctuations alone.  Lidz et al. used analytic and
numerical arguments to demonstrate that the scatter should rise
sharply as the mean opacity increases, leading to variations at $z
\sim 6$ on $\sim$40-50~\hinvMpc\ scales that are comparable to the
\citet{fan2006b} measurements.  If correct, this would significantly
weaken the direct evidence that the evolution in the UVB near $z \sim
6$ is related to patchy reionization.  Furthermore,
  \cite{bolton2007} and \cite{mesinger2009} demonstrated that even in
  the presence of a fluctuating UVB with a spatially invariant mean
  free path, the impact of the resulting ionization fluctuations on
  the effective optical depth is modest.  The largest fluctuations in
  the UVB typically occur in overdense regions which are already
  optically thick to \lya\ photons.  Any observational evidence for
  scatter in the \lya\ forest opacity in excess of that expected from
  density fluctuations or simple fluctuating UVB models alone would
  therefore be indicative of variations in the mean free path and
  spatial inhomogeneity in the IGM neutral fraction, which are potential hallmarks of reionization.

In this paper we provide a new analysis of the intergalactic
\lya\ opacity over $4 \lesssim z \lesssim 6$.  Our work is largely
motivated by deep Very Large Telescope (VLT)/X-Shooter observations of a single $z_{\rm em}
\sim 6$ quasar, \ulas0148\ ($z_{\rm em} = 5.98$), which was 
discovered in the UKIDSS Large Area Survey \citep{lawrence2007}.  As
we demonstrate below, this object shows an extremely dark ($\tau_{\rm
  eff} \gtrsim 7$) and extended ($\Delta l \simeq 110$~ \hinvMpc)
\lya\ trough.  Most remarkably, the trough extends down to $z \simeq
5.5$, where other lines of sight show high levels of transmitted flux.
We add \lya\ opacity measurements from \ulas0148\ and six other
$z_{\rm em} > 5.7$ quasars to the \citet{fan2006b} sample, along with
16 quasars over $4.5 \le z_{\rm em} \le 5.4$ observed at
moderate-to-high resolution to provide a lower-redshift sample for
comparison.  We compare measurements from this expanded sample to
predictions from simple IGM \lya\ transmission models based on numerical
simulations to determine whether fluctuations in the UVB are present.
Our hydrodynamical simulations include a suite of large boxes ($l_{\rm
  box} = 25$-100 \hinvMpc) in order to allow us both to evaluate the
expected scatter in \lya\ opacity from large-scale structure alone, as
well as couple simple fluctuating UVB models directly to the
density field.

\begin{figure*}
   \centering
   \begin{minipage}{\textwidth}
   \begin{center}
   \includegraphics[width=0.7\textwidth]{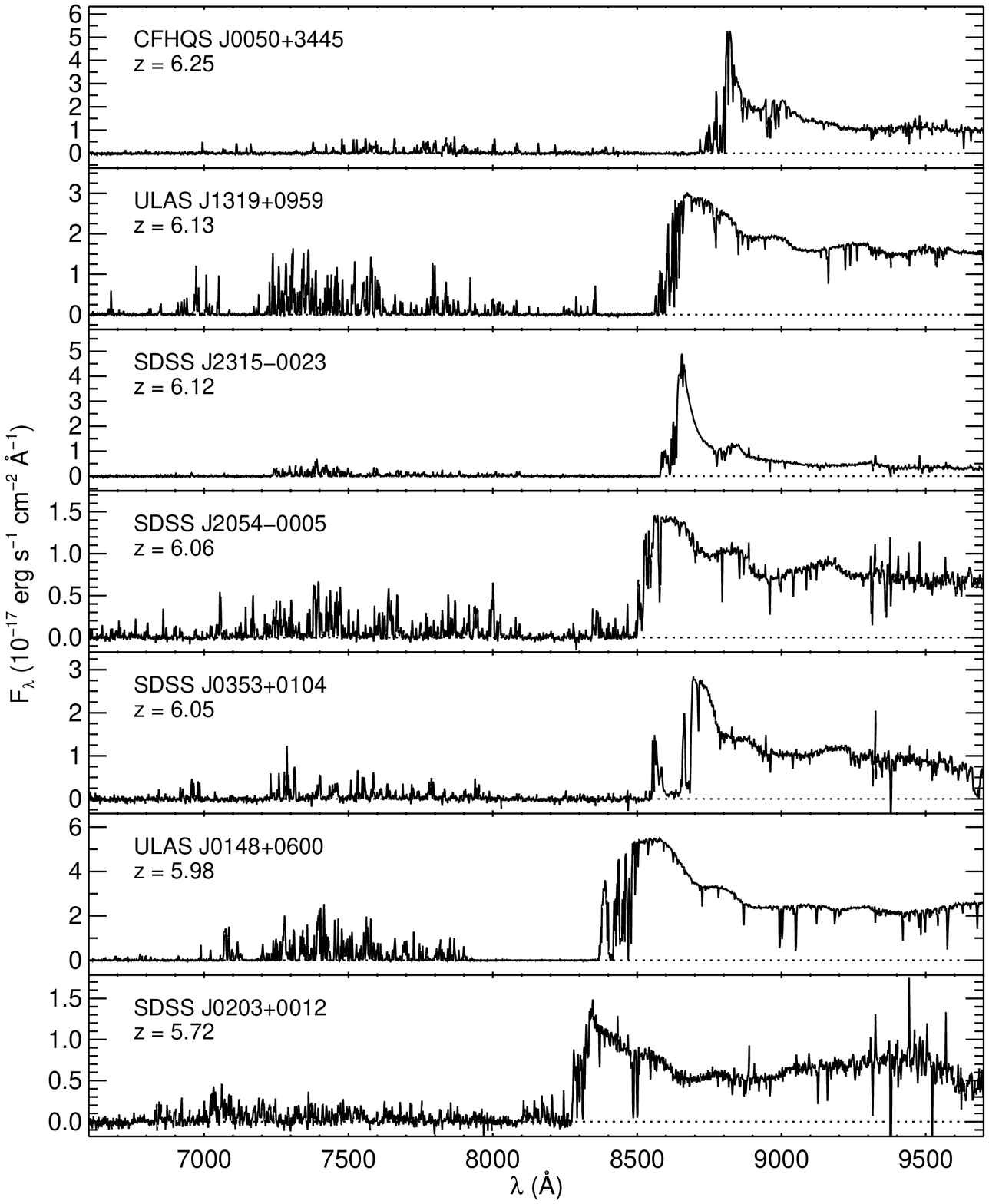}
   \vspace{-0.05in}
    \caption{Spectra of $z \sim 6$ quasars analysed in this work that are in addition to those in the \citet{fan2006b} sample.  ULAS~J1319$+$0950 and \ulas0148\ were observed with VLT/X-Shooter, while the remainder were observed with Keck/ESI (see Table~\ref{tab:qsos}).  Approximate fluxing is based on published $z'$-band magnitudes.  The spectra have been binned for display.  Note that the \lya\ forest flux for SDSS~J2315$-$0023 appears depressed because the $y$-axis has been scaled to accommodate the strong \lya\ emission line.}
   \label{fig:new_qsos}
   \end{center}
   \end{minipage}
\end{figure*}

We introduce the new data in Section~\ref{sec:data} and our numerical simulations in Section \ref{sec:simulations}.  In Section \ref{sec:models} we compare the \lya\ opacity measurements to predictions for a uniform UVB and for simple UVB models that assume the ionizing opacity is characterized by a single mean free path.  We argue that fluctuations in the mean free path must be present, and discuss the implications for the end of reionization in Section~\ref{sec:evolution}.  Our results are summarized in Section~\ref{sec:summary}.  Convergence tests for our models are presented in Appendix A.  We quote comoving distances generally assuming $(\Omega_{\rm m},
\Omega_{\Lambda}, h) = (0.308,
0.692, 0.678)$, consistent with recent
results from the {\it Planck} satellite \citep{planckXVI}.  Cosmological parameters are further discussed in Section~\ref{sec:simulations}.

\section{Data}\label{sec:data}

\subsection{Quasar Spectra}

This paper builds upon the sample of 19 $z_{\rm em} > 5.7$ quasars analysed by \citet{fan2006b} in two respects.  First, we add a further seven objects at $z_{\rm em} > 5.8$.  Notably, this new sample includes the $z_{\rm em} = 5.98$ quasar \ulas0148\footnote{For coordinates and magnitudes, see \citet{banados2014}.}, whose \lya\ trough is the primary motivation for this work.  Spectra for these objects are presented in Fig.~\ref{fig:new_qsos}.  We also add 16 quasars spanning $4.5 \le z_{\rm em} \le 5.4$, primarily to provide a lower-redshift baseline for evaluating the evolution of the \lya\ forest at $z > 5$.  Similar data at these redshifts were obtained by \citet{songaila2004}.  The present sample allows us to evaluate the evolution in \lya\ opacity, including its scatter between lines of sight, in a self-consistent manner over the entire redshift range $3.8 < z < 6.3$.  All spectra in this study were obtained at moderate or high spectral resolution with Keck/High Resolution Echelle Spectrograph (HIRES), Keck/Echellette Spectrograph and Imager (ESI), Magellan/Magellan Inamori Kyocera Echelle (MIKE), or VLT/X-Shooter, and thus are suited to the same type of analysis applied by \citet{fan2006b} to their Keck/ESI data.  A summary of the spectra is presented in Table~\ref{tab:qsos}.

\begin{table}
   \caption{List of QSOs analysed in this work that are in addition to those in the \citet{fan2006b} sample\vspace{-0.1in}}
   \label{tab:qsos}
   \begin{center}
   \begin{tabular*}{8.4cm}{@{\extracolsep{\fill}}lccc}
   \hline
    QSO  &  $z_{\rm em}$  &  Instrument  &  Ref.\textdagger \\
   \hline
   CFHQS~J0050$+$3445  &  6.25  &  ESI        &  5  \\ 
   ULAS~J1319$+$0950   &  6.13  &  X-Shooter  &  7  \\ 
   SDSS~J2315$-$0023   &  6.12  &  ESI	      &  5  \\
   SDSS~J2054$-$0005   &  6.06  &  ESI	      &  5  \\
   SDSS~J0353$+$0104   &  6.05  &  ESI	      &  5  \\
   ULAS~J0148$+$0600   &  5.98  &  X-Shooter  &  7  \\ 
   SDSS~J0203$+$0012   &  5.72  &  ESI	      &  5  \\
   SDSS~J0231$-$0728   &  5.42  &  X-Shooter  &  6  \\
   SDSS~J1659$+$2709   &  5.32  &  HIRES      &  3  \\
   SDSS~J1208$+$0010   &  5.27  &  X-Shooter  &  6  \\
   SDSS~J0915$+$4244   &  5.20  &  HIRES      &  2  \\
   SDSS~J1204$-$0021   &  5.09  &  HIRES      &  2  \\
   SDSS~J0040$-$0915   &  4.98  &  MIKE	      &  3  \\
   SDSS~J0011$+$1446   &  4.95  &  HIRES      &  3  \\
   SDSS~J2225$-$0014   &  4.89  &  MIKE	      &  4  \\
   SDSS~J1616$+$0501   &  4.88  &  MIKE	      &  4  \\
   BR~1202$-$0725      &  4.70  &  HIRES      &  1  \\
   SDSS~J2147$-$0838   &  4.60  &  MIKE       &  3  \\
   BR~0353$-$3820      &  4.59  &  MIKE       &  3  \\
   BR~1033$-$0327      &  4.52  &  MIKE	      &  7  \\
   BR~0006$-$6208      &  4.52  &  MIKE	      &  7  \\
   BR~0714$-$6449      &  4.49  &  MIKE       &  3  \\
   BR~0418$-$5723      &  4.48  &  MIKE       &  3  \\
   \hline
   \end{tabular*}
   \begin{flushleft}
   \textdagger~Spectrum references: 1-- \citet{lu1996}, 2 -- \citet{becker2006},
      3 -- \citet{becker2011a}, 4 -- \citet{calverley2011}, 5 -- \citet{becker2011b}, 
      6 -- \citet{becker2012}, 7 -- this work
   \end{flushleft}
   \end{center}
\end{table}

Reduction and continuum fitting procedures for all but two of our objects have been presented elsewhere (see Table~\ref{tab:qsos}).  For the $z_{\rm em} > 5.7$ objects, the continuum over the \lya\ forest was generally estimated using a power law normalized in regions relatively free of emission lines over $\sim$1285-1350~\AA\ in the rest frame, and out to $\sim$1450~\AA\ when possible.  A low-order spline fit was generally used for lower-redshift quasars, although the spline was typically placed near the power-law estimate.  Uncertainties in the \lya\ opacity measurements related to continuum fitting are discussed below.

New observations for \ulas0148\ and ULAS~J1319$+$0959 \citep{mortlock2009} were obtained with the X-Shooter spectrograph on the VLT \citep{sdodorico2006}.  Each object was observed for 10 h using 0.7 and 0.6 arcsec slits in the visible (VIS) and near-infrared (NIR) arms, respectively.  The spectra were flat-fielded, sky-subtracted using the method described by \cite{kelson2003}, optimally extracted \citep{horne1986} using 10~\kms\ bins, and corrected for telluric absorption using a suite of custom routines (see \citealt{becker2012} for more details).  These data will be described more fully in an upcoming work (Codoreanu et al., in preparation).  For \ulas0148\ we adopt a redshift of $z_{\rm em} = 5.98 \pm 0.01$ based on the peak of the \mgii\ emission line.  For ULAS~J1319$+$0959 we adopt $z_{\rm em} = 6.133$ based on [\cii] 158 $\mu$m measurements from \citet{wang2013}.

As discussed below, \ulas0148\ displays an extremely dark absorption trough in the \lya\ forest.  Since estimates of the mean opacity in such regions are sensitive to flux zero-point uncertainties, we adopted a reduction strategy intended to minimize such errors.  Individual exposures were combined using an inverse variance weighting scheme, where the variance in each two-dimensional reduced frame was estimated from the measured scatter about the sky model in regions not covered by the object trace, rather than derived formally from the sky model and detector characteristics.  This avoids biases when combining multiple exposures due to random errors in the sky estimate, which can be problematic when the sky background is relatively low.  We checked our combined one-dimensional X-shooter spectra for evidence of zero-point errors blueward of the quasar's Lyman limit, where there should be no flux, and found the errors to be negligible.

\subsection{\lya\ opacity measurements}

Following \citet{fan2006b}, we measure the mean opacity of the IGM to \lya\ in discreet regions along the lines of sight towards individual objects.  We quantify the opacity in terms of an effective optical depth, which is conventionally defined as $\tau_{\rm eff} = -\log{\langle F \rangle}$, where $F$ is the continuum-normalized flux.  Since our sample spans a broad redshift range, we measure \taueff\ in bins of fixed comoving length (50~\hinvMpc), rather than fixed redshift intervals.  This length scale, however, roughly matches the $\Delta z = 0.15$ bins used by \citet{fan2006b} over $z \sim 5-6$.

Our \lya\ flux measurements for all 23 objects are given in Table~\ref{tab:meanflux}.  Error estimates do not include continuum errors, which are instead incorporated into the modelling (see Section~\ref{sec:models}).  In order to avoid contamination from the quasar proximity region or from associated \lyb\ or \ovi\ absorption, we generally restrict our measurements to the region between rest-frame wavelengths 1041~\AA\ and 1176~\AA.  This also minimizes uncertainties in the continuum related to the blue wing of the \lya\ emission line.  For four of the six $z_{\rm em} > 5.9$ objects, however, we choose the maximum wavelength to lie just blueward of the apparent enhanced transmission in the proximity zone, as done by \citet{fan2006b}.  Exceptions to this are SDSS~J0353$+$0104, which is a broad absorption line (BAL) object, and SDSS~J2054$-$0005, for which edge of the region of enhanced flux is unclear.  In these cases we use a maximum rest-frame wavelength of 1176~\AA.  
 
\begin{figure}
   \begin{center}
   \includegraphics[width=0.45\textwidth]{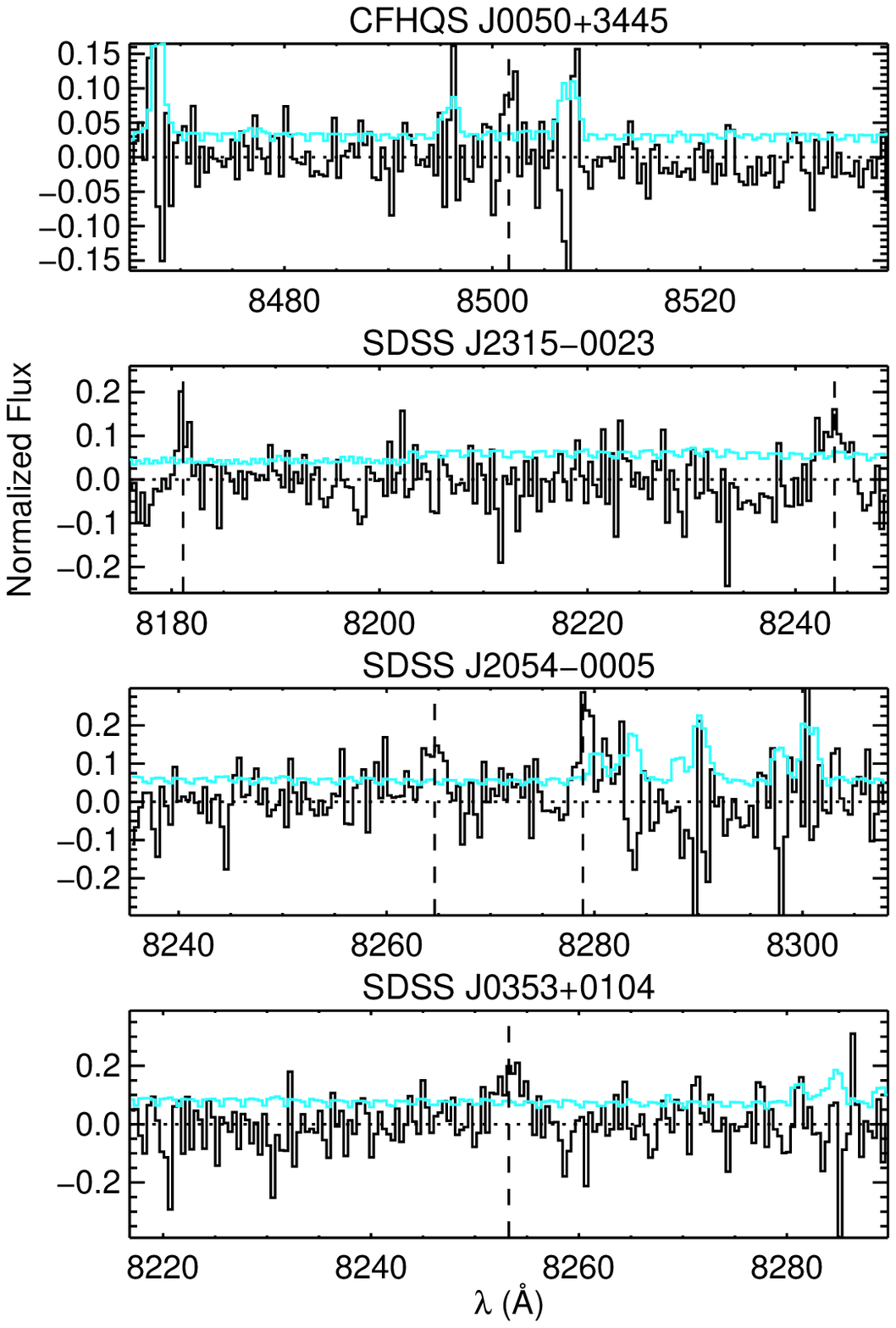}
   \vspace{-0.05in}
   \caption{Probable \lya\ transmission peaks in regions of the forest near $z \sim 6$ without a formal mean transmitted flux detection.  The dark histograms in each panel show the flux, while the light histograms show the 1$\sigma$ uncertainty.  The locations of probable transmission peaks are indicated by vertical dashed lines.}
   \label{fig:lya_peaks}
   \end{center}
\end{figure}

\begin{table}
   \caption{Formal continuum-normalized mean \lya\ transmitted flux
            measurements in 50~\hinvMpc\ regions.  The mean redshift in each section is given by $z_{\rm abs}$.  \vspace{-0.1in}}
   \label{tab:meanflux}
   \begin{center}
   \begin{tabular*}{8.4cm}{@{\extracolsep{\fill}}lccc}
   \hline
    QSO  &  $z_{\rm em}$  &  $z_{\rm abs}$  &  $\langle F \rangle$  \\
   \hline
CFHQS~J0050$+$3445  &  6.25  &  6.074  &  $   -0.00761 \pm 0.00196$  \\
                    &        &  5.902  &  $~~\,0.01149 \pm 0.00263$  \\
                    &        &  5.737  &  $~~\,0.01578 \pm 0.00209$  \\
                    &        &  5.577  &  $~~\,0.03836 \pm 0.00187$  \\
                    &        &  5.423  &  $~~\,0.10284 \pm 0.00178$  \\
ULAS1319+0950   &  6.13  &  5.948  &  $~~\,0.00820 \pm 0.00075$  \\
                &        &  5.781  &  $~~\,0.01404 \pm 0.00056$  \\
                &        &  5.620  &  $~~\,0.02544 \pm 0.00062$  \\
                &        &  5.464  &  $~~\,0.07504 \pm 0.00067$  \\
SDSS~J2315$-$0023   &  6.12  &  5.965  &  $   -0.00913 \pm 0.00273$  \\
                    &        &  5.797  &  $   -0.00094 \pm 0.00307$  \\
                    &        &  5.635  &  $~~\,0.03103 \pm 0.00197$  \\
                    &        &  5.479  &  $~~\,0.02652 \pm 0.00206$  \\
                    &        &  5.328  &  $~~\,0.05167 \pm 0.00150$  \\
                    &        &  5.183  &  $~~\,0.07156 \pm 0.00136$  \\
SDSS~J2054$-$0005   &  6.06  &  5.747  &  $~~\,0.00375 \pm 0.00280$  \\
                    &        &  5.586  &  $~~\,0.09962 \pm 0.00253$  \\
                    &        &  5.432  &  $~~\,0.07989 \pm 0.00227$  \\
                    &        &  5.283  &  $~~\,0.09959 \pm 0.00190$  \\
                    &        &  5.139  &  $~~\,0.16812 \pm 0.00196$  \\
SDSS~J0353$+$0104   &  6.05  &  5.737  &  $   -0.00561 \pm 0.00312$  \\
                    &        &  5.577  &  $~~\,0.02299 \pm 0.00281$  \\
                    &        &  5.423  &  $~~\,0.04055 \pm 0.00270$  \\
                    &        &  5.274  &  $~~\,0.06021 \pm 0.00210$  \\
                    &        &  5.130  &  $~~\,0.06568 \pm 0.00249$  \\
ULAS~J0148$+$0600   &  5.98  &  5.796  &  $~~\,0.00041 \pm 0.00037$  \\
                    &        &  5.634  &  $   -0.00013 \pm 0.00051$  \\
                    &        &  5.478  &  $~~\,0.04605 \pm 0.00043$  \\
                    &        &  5.327  &  $~~\,0.06339 \pm 0.00043$  \\
                    &        &  5.182  &  $~~\,0.13982 \pm 0.00040$  \\
SDSS0203+0012   &  5.72  &  5.423  &  $~~\,0.04361 \pm 0.00674$  \\
                &        &  5.275  &  $~~\,0.06883 \pm 0.00505$  \\
                &        &  5.131  &  $~~\,0.11174 \pm 0.00586$  \\
                &        &  4.992  &  $~~\,0.07700 \pm 0.00758$  \\
                &        &  4.858  &  $~~\,0.14021 \pm 0.00519$  \\
SDSS~J0231$-$0728   &  5.42  &  5.138  &  $~~\,0.23508 \pm 0.00062$  \\
                    &        &  4.999  &  $~~\,0.16729 \pm 0.00069$  \\
                    &        &  4.864  &  $~~\,0.17614 \pm 0.00063$  \\
                    &        &  4.734  &  $~~\,0.19190 \pm 0.00071$  \\
                    &        &  4.608  &  $~~\,0.19291 \pm 0.00074$  \\
SDSS~J1659$+$2709   &  5.32  &  5.043  &  $~~\,0.12661 \pm 0.00058$  \\
                    &        &  4.907  &  $~~\,0.16273 \pm 0.00059$  \\
                    &        &  4.776  &  $~~\,0.15605 \pm 0.00059$  \\
                    &        &  4.648  &  $~~\,0.25978 \pm 0.00068$  \\
                    &        &  4.525  &  $~~\,0.31753 \pm 0.00059$  \\
SDSS~J1208$+$0010   &  5.27  &  4.996  &  $~~\,0.25503 \pm 0.00107$  \\
                    &        &  4.861  &  $~~\,0.12560 \pm 0.00083$  \\
                    &        &  4.731  &  $~~\,0.22512 \pm 0.00100$  \\
                    &        &  4.605  &  $~~\,0.30364 \pm 0.00101$  \\
                    &        &  4.484  &  $~~\,0.29008 \pm 0.00105$  \\
SDSS~J0915$+$4244   &  5.20  &  4.929  &  $~~\,0.17196 \pm 0.00120$  \\
                    &        &  4.797  &  $~~\,0.10740 \pm 0.00103$  \\
                    &        &  4.669  &  $~~\,0.18277 \pm 0.00124$  \\
                    &        &  4.545  &  $~~\,0.16960 \pm 0.00132$  \\
                    &        &  4.425  &  $~~\,0.37799 \pm 0.00109$  \\
SDSS~J1204$-$0021   &  5.09  &  4.824  &  $~~\,0.19396 \pm 0.00147$  \\
                    &        &  4.696  &  $~~\,0.18355 \pm 0.00157$  \\
                    &        &  4.571  &  $~~\,0.35492 \pm 0.00172$  \\
                    &        &  4.450  &  $~~\,0.28457 \pm 0.00161$  \\
                    &        &  4.333  &  $~~\,0.32831 \pm 0.00153$  \\
SDSS~J0040$-$0915   &  4.98  &  4.720  &  $~~\,0.15083 \pm 0.00085$  \\
                    &        &  4.594  &  $~~\,0.28959 \pm 0.00088$  \\
                    &        &  4.473  &  $~~\,0.28650 \pm 0.00085$  \\
   \end{tabular*}
   \end{center}
\end{table}

\addtocounter{table}{-1}

\begin{table}
   \caption{-- {\it continued} \vspace{-0.1in}}
   \begin{center}
   \begin{tabular*}{8.4cm}{@{\extracolsep{\fill}}lccc}
   \hline
    QSO  &  $z_{\rm em}$  &  $z_{\rm abs}$  &  $\langle F/F_{\rm C} \rangle$  \\
   \hline
                    &        &  4.355  &  $~~\,0.35385 \pm 0.00095$  \\
                    &        &  4.242  &  $~~\,0.31313 \pm 0.00096$  \\
SDSS~J0011$+$1446   &  4.95  &  4.691  &  $~~\,0.29507 \pm 0.00045$  \\
                    &        &  4.567  &  $~~\,0.23270 \pm 0.00040$  \\
                    &        &  4.446  &  $~~\,0.42480 \pm 0.00037$  \\
                    &        &  4.329  &  $~~\,0.38612 \pm 0.00036$  \\
                    &        &  4.216  &  $~~\,0.34395 \pm 0.00036$  \\
SDSS~J2225$-$0014   &  4.89  &  4.634  &  $~~\,0.27845 \pm 0.00190$  \\
                    &        &  4.511  &  $~~\,0.29665 \pm 0.00181$  \\
                    &        &  4.393  &  $~~\,0.22136 \pm 0.00196$  \\
                    &        &  4.278  &  $~~\,0.29759 \pm 0.00206$  \\
                    &        &  4.166  &  $~~\,0.34961 \pm 0.00235$  \\
SDSS~J1616$+$0501   &  4.88  &  4.625  &  $~~\,0.20429 \pm 0.00184$  \\
                    &        &  4.502  &  $~~\,0.20725 \pm 0.00185$  \\
                    &        &  4.384  &  $~~\,0.45577 \pm 0.00221$  \\
                    &        &  4.269  &  $~~\,0.33738 \pm 0.00222$  \\
                    &        &  4.158  &  $~~\,0.29428 \pm 0.00254$  \\
BR~1202$-$0725      &  4.70  &  4.453  &  $~~\,0.27003 \pm 0.00090$  \\
                    &        &  4.336  &  $~~\,0.32692 \pm 0.00205$  \\
                    &        &  4.223  &  $~~\,0.40195 \pm 0.00216$  \\
                    &        &  4.113  &  $~~\,0.39365 \pm 0.00197$  \\
                    &        &  4.007  &  $~~\,0.47038 \pm 0.00191$  \\
SDSS~J2147$-$0838   &  4.60  &  4.358  &  $~~\,0.39491 \pm 0.00075$  \\
                    &        &  4.244  &  $~~\,0.42531 \pm 0.00087$  \\
                    &        &  4.134  &  $~~\,0.35357 \pm 0.00090$  \\
                    &        &  4.027  &  $~~\,0.34943 \pm 0.00091$  \\
                    &        &  3.923  &  $~~\,0.38447 \pm 0.00097$  \\
BR~0353$-$3820      &  4.59  &  4.349  &  $~~\,0.45173 \pm 0.00042$  \\
                    &        &  4.235  &  $~~\,0.28621 \pm 0.00043$  \\
                    &        &  4.125  &  $~~\,0.46655 \pm 0.00051$  \\
                    &        &  4.018  &  $~~\,0.27414 \pm 0.00045$  \\
                    &        &  3.915  &  $~~\,0.45430 \pm 0.00053$  \\
BR~1033$-$0327      &  4.52  &  4.282  &  $~~\,0.38280 \pm 0.00171$  \\
                    &        &  4.170  &  $~~\,0.27687 \pm 0.00179$  \\
                    &        &  4.062  &  $~~\,0.32834 \pm 0.00184$  \\
                    &        &  3.958  &  $~~\,0.45227 \pm 0.00200$  \\
                    &        &  3.856  &  $~~\,0.51813 \pm 0.00205$  \\
BR~0006$-$6208      &  4.52  &  4.282  &  $~~\,0.46323 \pm 0.00244$  \\
                    &        &  4.170  &  $~~\,0.38739 \pm 0.00287$  \\
                    &        &  4.062  &  $~~\,0.42680 \pm 0.00301$  \\
                    &        &  3.958  &  $~~\,0.46304 \pm 0.00323$  \\
                    &        &  3.856  &  $~~\,0.41029 \pm 0.00329$  \\
BR~0714$-$6449      &  4.49  &  4.253  &  $~~\,0.47062 \pm 0.00093$  \\
                    &        &  4.143  &  $~~\,0.29638 \pm 0.00104$  \\
                    &        &  4.035  &  $~~\,0.41984 \pm 0.00111$  \\
                    &        &  3.932  &  $~~\,0.40853 \pm 0.00116$  \\
                    &        &  3.831  &  $~~\,0.48802 \pm 0.00127$  \\
BR~0418$-$5723      &  4.48  &  4.244  &  $~~\,0.29407 \pm 0.00070$  \\
                    &        &  4.134  &  $~~\,0.41240 \pm 0.00075$  \\
                    &        &  4.027  &  $~~\,0.38504 \pm 0.00073$  \\
                    &        &  3.923  &  $~~\,0.45380 \pm 0.00075$  \\
                    &        &  3.822  &  $~~\,0.47881 \pm 0.00079$  \\
   \hline
   \end{tabular*}
   \end{center}
\end{table}

Where no transmitted flux is formally detected, we adopt a lower limit on \taueff\ assuming a mean transmitted flux equal to twice the formal uncertainty.   In these cases, we also searched the spectra for individual transmission peaks whose flux may have been smaller than the formal uncertainty for the total 50~\hinvMpc\ region.  A peak was considered significant if it had at least four adjacent pixels that exceeded the 1$\sigma$ error estimate, and if the combined significance of the flux in these pixels was $\ge$5$\sigma$.  The identified peaks are shown in Fig.~\ref{fig:lya_peaks}.  In regions where one or more peaks were detected, we adopt an upper limit on \taueff\ assuming that the total flux in that 50~\hinvMpc\ region is equal to the 2$\sigma$ lower limit on the flux in those peaks alone.  

\begin{table}
   \caption{Lower limits on the continuum-normalized \lya\ flux for 50~\hinvMpc\ regions
            that do not have a formal 2$\sigma$ detection yet show individually significant transmission peaks\vspace{-0.1in}}
   \label{tab:limits}
   \begin{center}
   \begin{tabular*}{8.4cm}{@{\extracolsep{\fill}}lccc}
   \hline
    QSO  &  $z_{\rm em}$  &  $z_{\rm abs}$  &  $\langle F/F_{\rm C} \rangle$  \\
   \hline
CFHQS~J0050$+$3445  &  6.25  &  6.074  &  $>$0.0005  \\
SDSS~J2315$-$0023   &  6.12  &  5.797  &  $>$0.0021  \\
SDSS~J2054$-$0005   &  6.06  &  5.747  &  $>$0.0010  \\
SDSS~J0353$+$0104   &  6.05  &  5.737  &  $>$0.0018  \\
   \hline
   \end{tabular*}
   \end{center}
\end{table}

\begin{figure*}
   \centering
   \begin{minipage}{\textwidth}
   \begin{center}
   \includegraphics[width=1.0\textwidth]{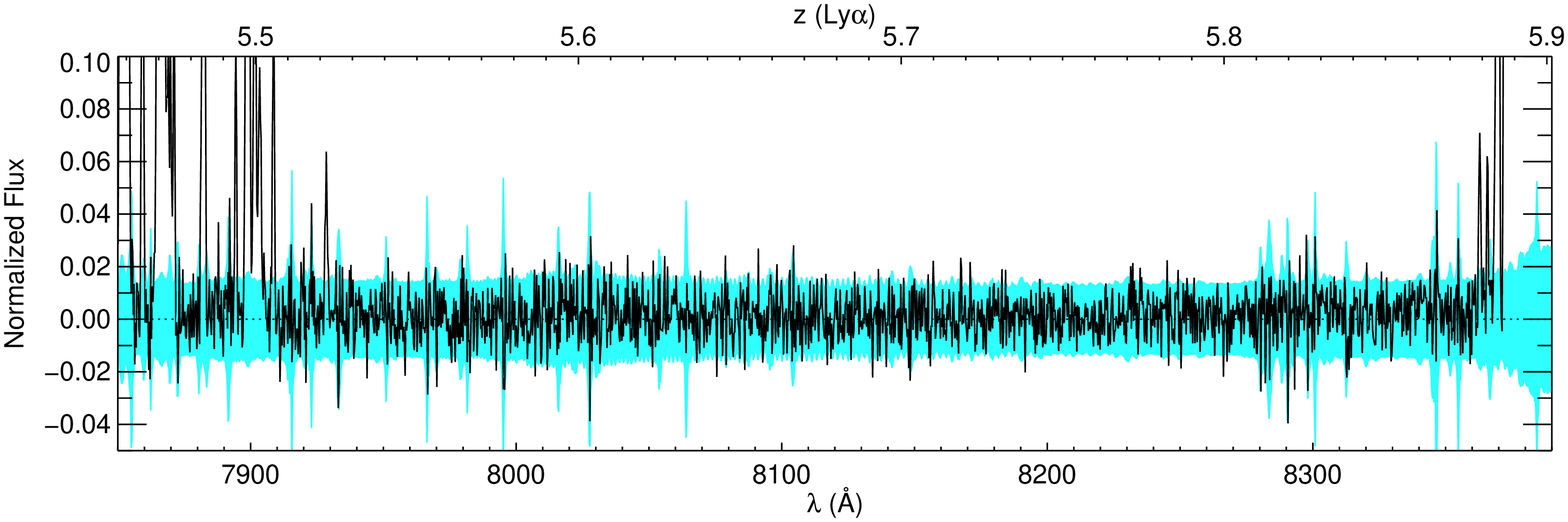}
   \vspace{-0.05in}
    \caption{The \lya\ trough towards \ulas0148.  The flux, binned in 10~\kms\ pixels, is shown by the dark histogram.  The shaded region shows the $\pm$2$\sigma$ uncertainty interval.  Note the scale on the vertical axis.  \lya\ redshifts are shown at the top of the plot.  No significant \lya\ transmission peaks appear over $5.523 \le z \le 5.879$.}
   \label{fig:trough}
   \end{center}
   \end{minipage}
\end{figure*}

Following \citet{fan2006b}, we made no attempt to correct for contamination from intervening metal lines or damped \lya\ systems (DLAs).  Metal lines, at least at $z < 5$, generally account for only a few percent of the absorption in the \lya\ forest \citep{schaye2003,kirkman2005,kim2007,becker2011a}, and so are not expected to strongly affect \taueff\ measurements at $z \gtrsim 4$ or add significantly to their scatter on 50~\hinvMpc scales.  DLAs are potentially more problematic, as a single system will increase \taueff\ in a 50~\hinvMpc\ region by $\sim$0.4.  They become increasingly difficult to identify at $z > 5$, however, where the high levels of absorption in the \lya\ forest mean that DLAs must often be identified via their associated metal lines, coverage of which varies between lines of sight and is often incomplete.  We tested the impact of DLAs on \taueff\ at $z < 5$ by repeating our measurements after masking DLAs visible in the \lya\ forest.  This naturally lowered \taueff\ in some regions, although the difference was not large enough to affect the interpretation of the data presented below.  We detect no metal lines in the spectrum of \ulas0148\ over the redshift range spanned by its \lya\ trough.  A more detailed inventory of metals along this line of sight will be presented by Codoreanu et al.  (in preparation).  Here we note, however, that a detailed search revealed no metal systems at $z > 5.5$ traced by \civ, \siiv, or \mgii\ doublets, or by groups of low-ionization lines such as \oi, \cii, \siii, \feii, and \alii.  We are therefore reasonably confident that classical metal-enriched DLAs are not affecting this line of sight\footnote{Our \oi\ column density detection limit over this redshift range is $N_{\rm O\,I} \simeq 10^{13.5}~{\rm cm^{-2}}$, which corresponds to a DLA ($N_{\rm H\,I} \ge 10^{20.3}~{\rm cm^{-2}}$) metallicity of ${\rm [O/H]} \lesssim -3.5$.  This is a factor of 5 in metallicity below the most metal-poor DLAs reported in the literature \cite[e.g.,][and references therein]{cooke2011b}.}, and so are not responsible for the most extreme values of \taueff\ in our data.  Future, more detailed analysis of the \lya\ forest at these redshifts, however, may require a more comprehensive treatment of both metal lines and optically thick absorbers.

\subsubsection{The \ulas0148\ \lya\ trough}

The \lya\ forest towards \ulas0148\ displays an unusually long ($\sim$110~\hinvMpc) \lya\ trough over $5.523 \le z \le 5.879$ (Fig.~\ref{fig:trough}).  This is roughly twice as long as the longest troughs reported previously \citep{white2003,fan2006b}, and extending down to significantly lower redshifts.  As discussed in more detail below, the depth of the trough towards \ulas0148\ is in significant contrast with other lines of sight at the same redshifts.  For the 50~\hinvMpc\ regions centred at $z=5.63$ and 5.80 we find 2$\sigma$ lower limit of $\tau_{\rm eff} \ge 6.9$ and 7.2, respectively.  For the complete trough we measure $\tau_{\rm eff} \ge 7.4$.  

\begin{figure}
   \begin{center}
   \includegraphics[width=0.45\textwidth]{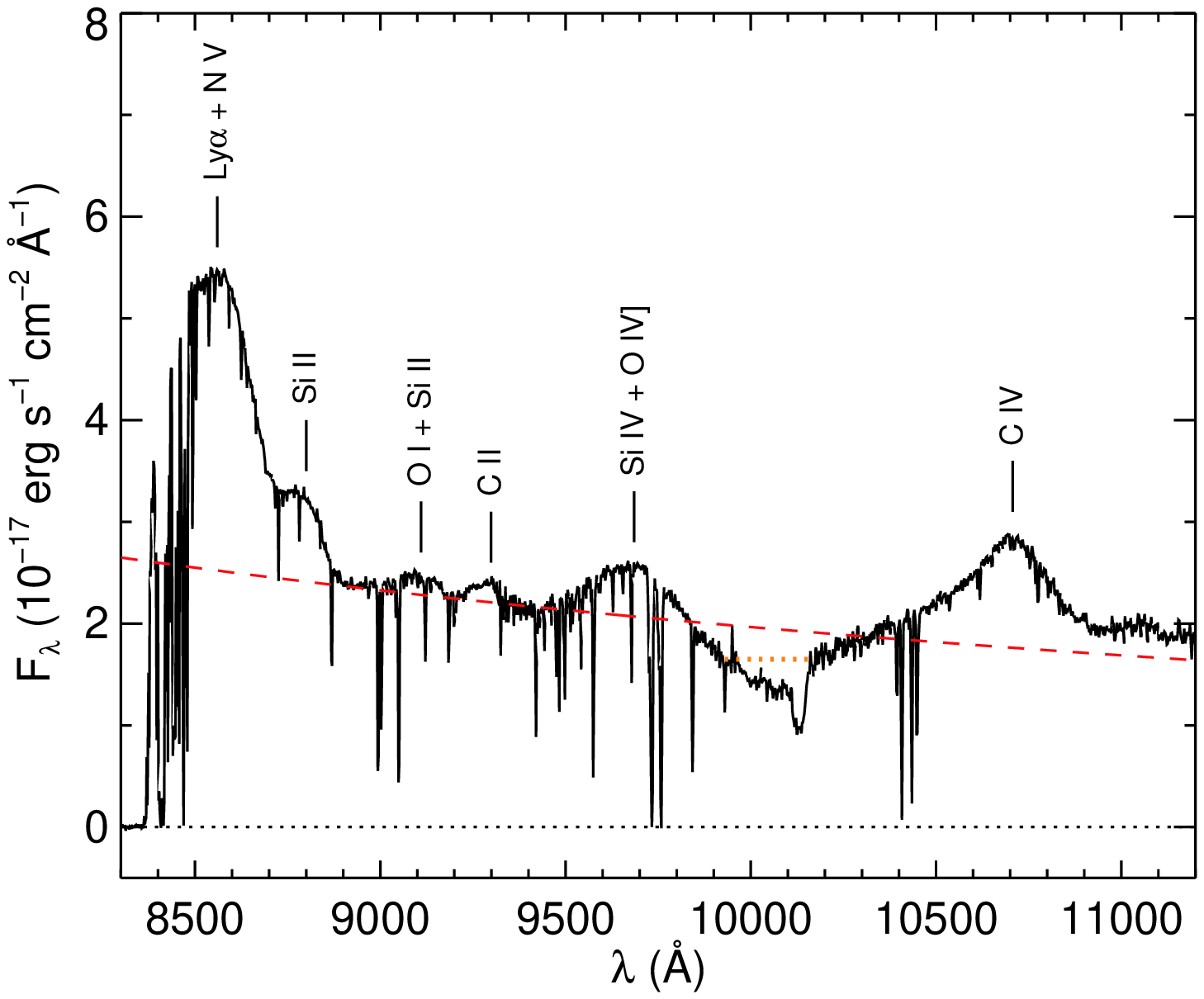}
   \vspace{-0.05in}
   \caption{X-Shooter spectrum of \ulas0148\ redward of the \lya\ forest.  Prominent emission lines are marked.  The dashed line is an estimate of the underlying power-law continuum.  The \civ\ BAL is visible as a depression below this estimate between the \siiv\ and \civ\ emission lines.  A mini-BAL component at $z \simeq 5.54$ is visible near 10130~\AA. }
   \label{fig:bal}
   \end{center}
\end{figure}

As noted above, we find no intervening metal absorbers over this redshift interval that would suggest the presence of DLAs.  \ulas0148\ does show a mild, broad depression in its spectrum between the \siiv\ and \civ\ emission lines ($9840\,{\rm \AA} \lesssim \lambda \lesssim 10320\,{\rm \AA}$) compared to a power-law estimate of the continuum, however (Fig.~\ref{fig:bal}).  BALs are therefore a potential concern, since a \civ\ BAL over this interval could indicate \lya\ and/or \nv\ absorption in the \lya\ forest.  The clearest indication of broad \civ\ absorption is the narrow mini-BAL feature over $10110\,{\rm \AA} \lesssim \lambda \lesssim 10150\,{\rm \AA}$.  It is unclear, however, the extent to which the remainder of the depression in Fig.~\ref{fig:bal} is due to broad absorption.  A study of BAL quasars in the Sloan Digital Sky Survey (SDSS) by \citet{allen2011} found no compelling examples of such wide, shallow BALs (P. Hewett, private communication).  The lack of distinct features (apart from the mini-BAL) between the \siiv\ and \civ\ emission lines instead suggests that most of the depression is an intrinsic feature of the quasar spectrum, rather than a BAL.  No broad absorption in \siiv\ is seen, which further suggests that  most of the depression is not a \civ\ BAL, although the gas could have either a high ionization state or a low column density. We suspect that the  genuine broad \civ\ absorption is restricted to at most a modest depression over $9930 \lesssim \lambda \lesssim 10170\,{\rm \AA}$, indicated by the dotted line in Fig.~\ref{fig:bal}.  To be conservative, however, we assume that the entire depression below the power-law estimate is due to a \civ\ BAL, and model the impact of corresponding \nv\ absorption.  We modify our continuum estimate for \ulas0148\ by including an estimate of the \nv\ broad absorption that assumes the \civ\ and \nv\ BAL profiles are similar in velocity structure and amplitude \citep[e.g.,][]{baskin2013}.  This has a relatively minor affect, decreasing \taueff\ in the 50~\hinvMpc\ region centred at $z = 5.63$ by 0.3 and by $<$0.1 elsewhere.  The flux measurements given in Table~\ref{tab:meanflux} and the \taueff\ lower limits quoted above take this estimate of the \nv\ absorption into account.  
The lack of \siiv\ suggests that any BAL would be weak in \lya\ (Hamann et al., in preparation).  The strongest potential \civ\ absorption, moreover, occurs at $z < 5.54$, which for \lya\ falls blueward of the trough.  We repeat our modelling procedure for \ovi~$\lambda$1032,1038 broad absorption that may be present in the \lyb\ forest, however.  

\begin{figure*}
   \centering
   \begin{minipage}{\textwidth}
   \begin{center}
   \includegraphics[width=1.0\textwidth]{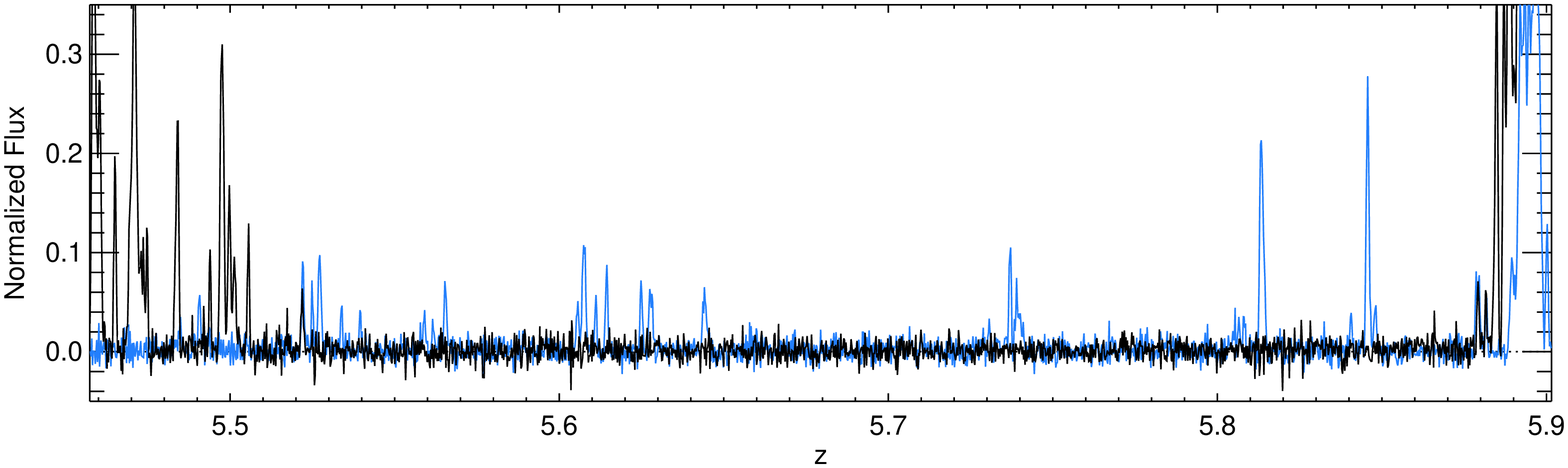}
   \vspace{-0.05in}
    \caption{The \lya\ trough (black) towards \ulas0148\ overlaid with the \lyb\ forest (blue) at the same redshifts.  \lyg\ absorption also occurs in the \lyb\ forest at $z \le 5.63$.}
   \label{fig:trough_with_lyb}
   \end{center}
   \end{minipage}
\end{figure*}

Although no \lya\ transmission peaks are detected in the
\ulas0148\ trough, we can set an upper limit on the \lya\ opacity
using the fact that transmission is seen in the \lyb\ forest over the
same redshifts (Fig.~\ref{fig:trough_with_lyb}).  We use
hydrodynamical simulations (see Section~\ref{sec:simulations})
to model the transmission in the \lyb\ forest over the
$\sim$80~\hinvMpc\ interval from $z = 5.62$ to 5.88 ($6790 \lesssim
\lambda \lesssim 7060$), where the upper limit in redshift corresponds
to the start of the \lya\ trough, and the lower limit is set by the
onset of \lyg\ absorption for a quasar at redshift $z_{\rm em} =
5.98$.  Over this interval we measure a total effective optical depth
of $\tau_{\rm eff}^{\rm tot} = 5.17 \pm 0.05$.  We modeled this
absorption by superposing simulated \lyb\ forest spectra at $z_{\rm
  trough} = 5.620$ or 5.831 on to foreground \lya\ absorption at
$z_{\rm fg} = (1+z_{\rm trough})(\lambda_{\alpha}/\lambda_{\beta})-1$.
The \lyb\ and \lya\ spectra were drawn randomly, with the optical
depths in the foreground \lya\ sample collectively scaled to reproduce
the mean \lya\ opacity at $z_{\rm fg}$ measured by
\citet{becker2013a}.  For each trial, we then scaled the \lyb\ optical
depths such that the combined opacity matched our measured value in
the \lyb\ forest, and then calculated the corresponding \lya\ opacity
at $z=z_{\rm trough}$.  In principle this procedure can be used to set
both lower and upper bounds of \taueff\ for \lya; however, we find
that the conversion from \taueffb\ to \taueffa\ is not converged for
our simulations (see Appendix~\ref{app:lyb_convergence}), in the sense
that \taueffa\ is probably too high for a given \taueffb\ for even our
highest resolution simulation.  A lower limit on \taueffa\ in the
trough set by this procedure would therefore not be reliable, although
an upper limit will be conservative.  For the
$\sim$80~\hinvMpc\ stretch containing only \lyb\ and foreground
\lya\ absorption, our measurement of $\tau_{\rm eff}^{\rm tot}$
implies a 95 percent upper limit of $\tau_{\rm eff}^{\alpha} \le
12.3$.  Here we have interpolated the results from adopting $z_{\rm
  trough} = 5.620$ and 5.831 for our simulations on to the mean
redshift of the \lyb\ trough ($z \simeq 5.75$).

\begin{figure}
   \begin{center}
   \includegraphics[width=0.45\textwidth]{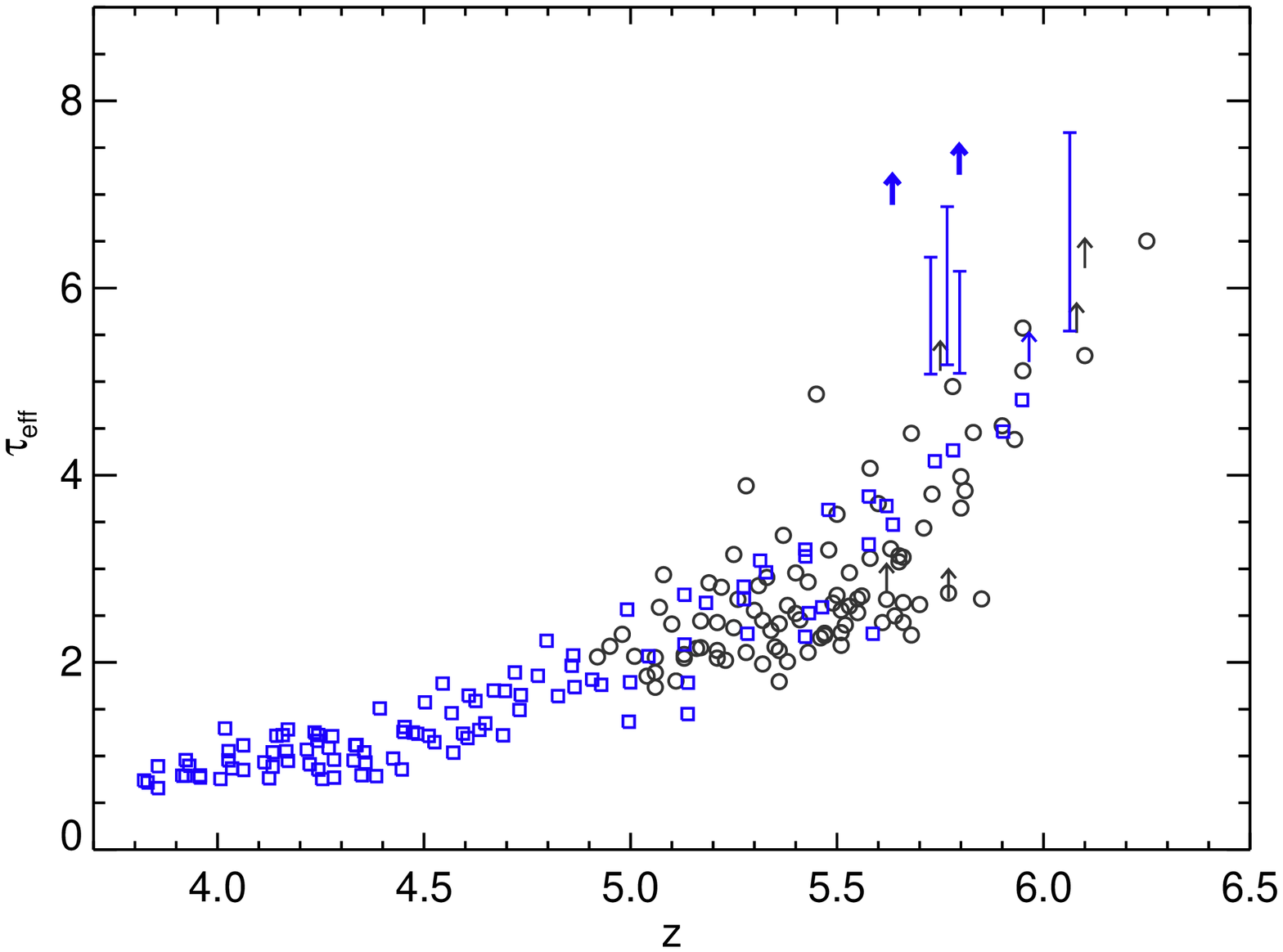}
   \vspace{-0.05in}
   \caption{\lya\ \taueff\ measurements.  Black circles and arrows (lower limits) are from \citet{fan2006b}.  Blue squares, error bars, and arrows are from this work.  The points at $z \simeq 5.63$ and 5.80 with $\tau_{\rm eff} \gtrsim 7$ are from the line of sight towards \ulas0148.}
   \label{fig:taueff_data}
   \end{center}
\end{figure}

The combined \lya\ \taueff\ data are shown in
Fig.~\ref{fig:taueff_data}.  As pointed out by \citet{fan2006b},
\taueff\ exhibits both a strong overall increase with redshift and an
enhanced scatter at $z > 5$.  Our new measurements support this trend,
with the \ulas0148\ trough providing the starkest demonstration that
lines of sight with very strong absorption exist at the same redshift
as lines of sight where the absorption is far more modest.  The
primary goal of this paper is to determine whether these
sightline-to-sightline variations are predicted by simple models
of the UVB, or whether more complicated effects--potentially relating
to hydrogen reionization--are needed.  We now turn to interpreting the
\taueff\ measurements within the context of simple models for the
evolution of the ionizing UVB.  These models jointly consider the
large-scale radiation and density fields using the numerical
simulations described below.

\section{Hydrodynamical simulations}\label{sec:simulations}

The large scale distribution of gas in the IGM at $z>4$ is modelled in
this work using a set of 11 cosmological hydrodynamical simulations.
These simulations are summarized in Table~\ref{tab:hydrosims}, and
were performed using the smoothed particle hydrodynamics code
{\sc gadget}-3, which is an updated version of the publicly available code
{\sc gadget}-2 last described by \cite{springel2005}.

The fiducial cosmological parameters adopted in the simulations are
$(\Omega_{\rm m}, \Omega_{\Lambda}, \Omega_{\rm b}h^2, h, \sigma_8,
n_{\rm s}) = (0.26, 0.74, 0.023, 0.72, 0.80, 0.96)$.  These
calculations were all started at redshift $z=99$, with initial
conditions generated using the \cite{eisensteinhu1999} transfer
function.  The gravitational softening length was set to 1/25th the
mean linear interparticle separation.  Star formation was modelled
using an approach designed to optimize \lya~forest simulations, where
all gas particles with overdensity $\Delta=\rho/\langle \rho \rangle
>10^{3}$ and temperature $T<10^{5}\rm\,K$ are converted into
collisionless star particles.  The photo-ionization and heating of the
IGM were included using a spatially uniform UVB, applied
assuming the gas in the simulations is optically thin
(\citealt{hm2001}).  The fiducial thermal history in this work
corresponds to model C15 described in \cite{becker2011a}; see also
Appendix~\ref{app:num_convergence}.

A total of nine simulations were performed to test the impact of box size
and resolution on our results (although the two models we use most
extensively in this work are the 100--1024 and 25--1024 simulations
listed in Table~\ref{tab:hydrosims}).  These span a range of box sizes
and gas particles masses, from 25 to 100~\hinvMpc\ and 1.79 to $7.34 \times
10^{8} {\rm\,M_{\odot}}h^{-1}$.  Note, however, that these models
(particularly 100--1024) employ a rather low mass resolution relative
to that required for fully resolving the low density \lya\ forest at
$z>5$ (cf. \citealt{boltonbecker2009}, who recommend $L\geq
40$~\hinvMpc\ and $M_{\rm gas}\leq 2\times 10^{5}
{\rm\,M_{\odot}}h^{-1}$).

In this work, however, our goal is to examine spatial fluctuations in
the \lya~forest opacity and UVB on large scales.  The
typical scales are difficult to capture correctly in smaller ($\sim$10~\hinvMpc) boxes with high mass resolution; the mean free path
to Lyman limit photons at $z=5$ is $\sim$60~\hinvMpc\ 
\citep[comoving; e.g.,][]{prochaska2009b,songaila2010,worseck2014}. Since
computational constraints mean we are unable to perform simulations in
boxes with $L\sim 100$~\hinvMpc\ at the mass resolution needed to
fully resolve the low density IGM, a compromise must then be made on
this numerical requirement.  We have, however, verified that this
choice will not alter the main conclusions of this study.  This is
examined in further detail in Appendix~\ref{app:num_convergence},
where we present a series of convergence tests with box size and mass
resolution.

In addition to the nine simulations used to test box size and mass
resolution convergence, we also perform two further simulations in
which the cosmological parameters and IGM thermal history are varied.
These models are used to test the impact of these assumptions on our
results. The Planck simulation adopts $(\Omega_{\rm m},
\Omega_{\Lambda}, \Omega_{\rm b}h^2, h, \sigma_8, n_{\rm s}) = (0.308,
0.692, 0.0222, 0.678, 0.829, 0.961)$, consistent with the recent
results from {\it Planck} \citep{planckXVI}.  The Dz12\_g1.0 model
adopts an alternative IGM heating history which reionizes earlier
($z_{\rm r} = 12$, cf. $z_{\rm r} = 9$ for our fiducial model), and
heats the gas in the low density IGM to higher temperatures. Further
details and tests using these models maybe found in
Appendices~\ref{app:cosmology} and \ref{app:Thistory}.

Finally, we extract synthetic \lya~forest spectra from the output of
the hydrodynamical simulations using a standard approach
(e.g. \citealt{theuns1998}) under the assumption of a spatially
uniform \hi~photo-ionization rate, $\Gamma_{\rm HI}$. As we now discuss
in the next section, these spectra will also be generated using a
model for spatial fluctuations in the ionization rate which is applied
in post-processing.

\begin{table}
  \centering

  \caption{Hydrodynamical simulations used in this work.  The columns
    list the model name, the box size in comoving $h^{-1}\rm\,Mpc$,
    the number of gas and dark matter particles, and the gas particle
    mass for each simulation.  The fiducial cosmological parameters
    adopted in the simulations are $(\Omega_{\rm m}, \Omega_{\Lambda},
    \Omega_{\rm b}h^2, h, \sigma_8, n_{\rm s}) = (0.26, 0.74, 0.023,
    0.72, 0.80, 0.96)$, and the fiducial UVB is from
    \citet{hm2001}.  The two exceptions are the Planck model, which
    adopts an alternative set of cosmological parameters, and the
    Dz12\_g1.0 model, which assumes an alternative thermal history
    (see text for further details).}
  \begin{tabular}{c|c|c|c}
    \hline
    Model & $L$ (\hinvMpc)  & Particles & $M_{\rm gas}$ $(M_{\odot}\,h^{-1})$\\
    \hline
    100--1024   & 100 & $2\times 1024^{3}$ & $1.15 \times 10^{7}$ \\ 
    100--512    & 100 & $2\times 512^{3}$  & $9.18 \times 10^{7}$ \\
    100--256    & 100 & $2\times 256^{3}$  & $7.34 \times 10^{8}$ \\
    50--1024    & 50  & $2\times 1024^{3}$ & $1.43 \times 10^{6}$ \\
    50--512     & 50  & $2\times 512^{3}$  & $1.15 \times 10^{7}$ \\
    50--256     & 50  & $2\times 256^{3}$  & $9.18 \times 10^{7}$ \\
    25--1024    & 25  & $2\times 1024^{3}$ & $1.79 \times 10^{5}$ \\
    25--512     & 25  & $2\times 512^{3}$  & $1.43 \times 10^{6}$ \\
    25--256     & 25  & $2\times 256^{3}$  & $1.15 \times 10^{7}$ \\
    Planck      & 100 & $2\times 1024^{3}$ & $1.25 \times 10^{7}$ \\
    Dz12\_g1.0  & 100 & $2\times 1024^{3}$ & $1.15 \times 10^{7}$ \\
  \hline
  \end{tabular}
\label{tab:hydrosims}
\end{table}



\section{UV Background Models}\label{sec:models}

\subsection{Uniform UVB}\label{sec:uniform_uvb}

We begin by considering models with a uniform ionizing background,
where the scatter in \taueff\ between lines of sight is driven
entirely by variations in the density field.  \citet{lidz2006} found
that such a model could potentially accommodate much of the observed
scatter in \taueff\ without invoking additional factors such as
fluctuations in the UVB related to the end of reionization.  Our
first task is therefore to re-assess this conclusion in light of the
additional data presented herein.


We calculate the expected scatter in \taueff\ at each simulation
redshift by fixing the volume-averaged neutral fraction, \fhi,
assuming a uniform UVB, and measuring the mean flux along randomly
drawn 50~\hinvMpc\ sections of \lya\ forest.  We use 100--1024
simulation for our fiducial estimates in order to include the maximum
amount of large-scale structure.  Trials with the other
  simulations in Table~\ref{tab:hydrosims} show relatively little
dependence on box size, but decreased scatter in \taueff\ towards
higher mass resolution (see Appendix~\ref{app:num_convergence}).  This
trend is driven by the fact that the low-density regions, which
dominate the transmission at these redshifts, become better resolved
with decreasing particle mass \citep[see][]{boltonbecker2009}.  Our
choice of the 100~\hinvMpc\ box is therefore conservative in
determining whether the observed scatter can be reproduced with a
uniform UVB.

Our nominal \fhi\ evolution is given by
\begin{equation}\label{eq:fHI}
   \langle f_{\rm H\, I} \rangle (z) = (1.3 \times 10^{-5}) \left( \frac{1+z}{5.6} \right)^{\eta} \, ,
\end{equation}
with $\eta = 2.9$ (5.0) for $z \le 4.6$ ($z > 4.6$).  The evolution at
$z \le 4.6$ is chosen to reproduce the mean opacity measurements of
\citet{becker2013a} (although we note that the precise neutral fraction will depend on the simulation parameters,
primarily mass resolution).  The evolution in \fhi\ at $z > 4.6$ is
chosen such that the lower bound in \taueff\ roughly traces the lower
envelope of the observed values over $4.6 < z < 5.8$.  At these
redshifts, the majority of \taueff\ measurements tend to cluster along
a relatively narrow locus bounded by this envelope, while outlying
points tend to scatter towards higher values.  While our choice of
\fhi\ evolution at $z > 4.6$ is not a fit, anchoring the
\taueff\ distribution near this lower boundary provides one way of
determining whether all of the data, at least up to $z \simeq 5.8$,
can be reproduced using a uniform UVB.

\begin{figure}
   \begin{center}
   \includegraphics[width=0.45\textwidth]{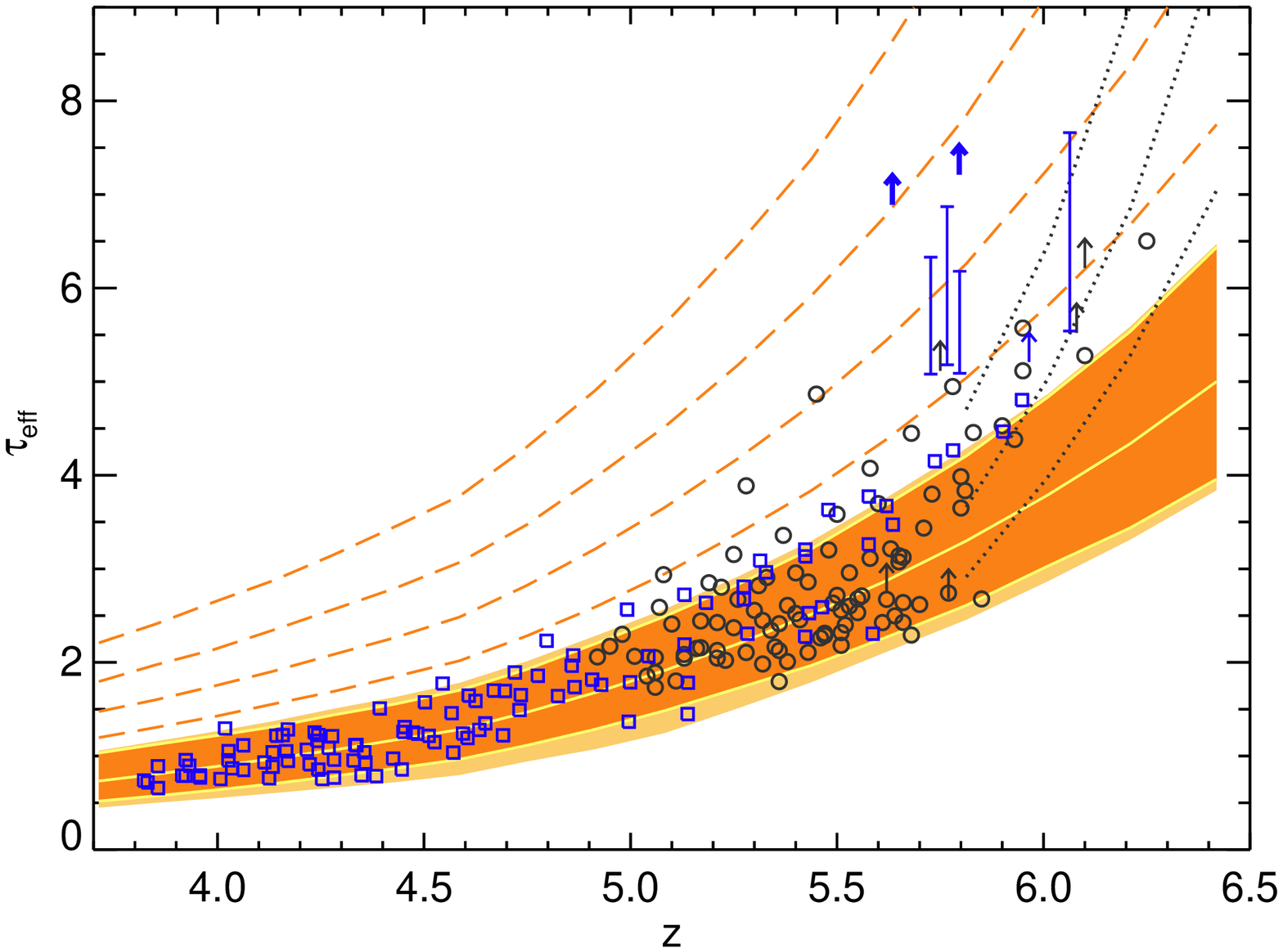}
   \vspace{-0.05in}
   \caption{Predicted distribution of \lya\ \taueff\ for our uniform UVB model.  Data points are as in Fig.~\ref{fig:taueff_data}.  The dark shaded region spans the two-sided 95 per cent range in \taueff\ for the evolution of the hydrogen neutral fraction given by equation~(\ref{eq:fHI}).  The light shaded bands on either side of this region show the additional scatter due to random continuum errors (see text).  The dashed lines give the one-sided 95 per cent upper limit in \taueff\ for 50~\hinvMpc\ regions when the neutral fraction is increased by 0.15, 0.3, 0.45, and 0.6 dex (bottom to top).  The dotted lines give the two-sided 95 per cent interval in \taueff\ for a separate evolution in the neutral fraction (see text).}
   \label{fig:taueff_uniform}
   \end{center}
\end{figure}

\begin{figure}
   \begin{center}
   \includegraphics[width=0.45\textwidth]{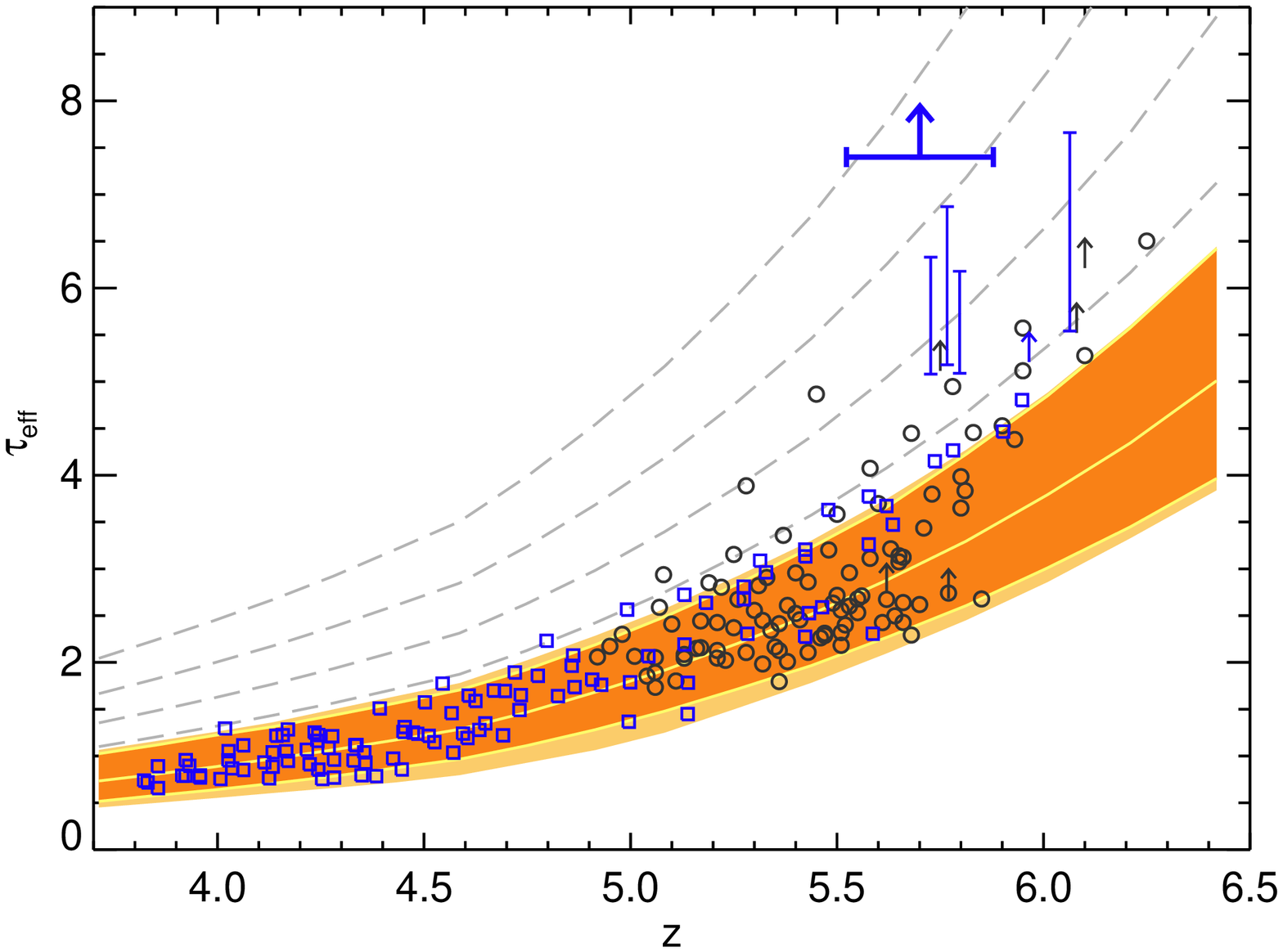}
   \vspace{-0.05in}
   \caption{Same as Fig.~\ref{fig:taueff_uniform}, except that the 50~\hinvMpc\ regions at $z = 5.63$ and 5.80 towards \ulas0148\ have been replaced by the lower limit for the complete \lya\ trough (large arrow).   The dashed lines give the one-sided 95 per cent upper limit in \taueff\ for 100~\hinvMpc\ regions when the neutral fraction is increased by 0.15, 0.3, 0.45 and 0.6 dex (bottom to top) versus the nominal value at each redshift.}
   \label{fig:taueff_uniform_trough}
   \end{center}
\end{figure}

Our uniform UVB model is compared to the \taueff\ measurements in
Fig.~\ref{fig:taueff_uniform}.  As noted above, we do not include
continuum uncertainties in the \taueff\ values measured from the data,
but instead incorporate these effects directly into our models.  The
dark shaded region in Fig.~\ref{fig:taueff_uniform} shows the
predicted range in \taueff\ without continuum errors, while the outer,
lighter shaded regions includes random continuum errors with an rms amplitude linearly interpolated between 5, 10, and 20 per
cent at $z=3$, 4 and 5, respectively, and a constant 20 per cent at $z
> 5$.  At $z < 5$, the scatter in the data is well reproduced by the
simulations, suggesting that the UVB near 1 Ryd is reasonably uniform
at these redshifts.  This is not surprising given that the mean free
path to hydrogen ionizing photons at $z < 5$ is long with respect to the typical separation between star-storming galaxies 
\citep[e.g.,][]{prochaska2009b,songaila2010,worseck2014}.  Even up to
$z \lesssim 5.3$ the scatter in \taueff\ outside that expected for a
uniform UVB is minimal.

Over $5.3 < z < 5.8$ the scatter in the uniform UVB model still spans a large fraction of the data; however, an increasing number of points fall above the upper model bound with increasing redshifts.  The most extreme scatter occurs near $z \simeq 5.6$ where, in order to span the collection of points with $\tau_{\rm eff} \simeq 2.5$, the 97.5 per cent upper limit for the uniform UVB model is $\tau_{\rm eff} \le 3.7$.  In contrast, the 50~\hinvMpc\ section at $z=5.63$ towards \ulas0148\ has $\tau_{\rm eff} > 6.9$.  Several other points, although not as extreme as the \ulas0148\ values, also lie significantly above the upper bound in \taueff, even down to $z \simeq 5.3$.  This strongly suggests that the \lya\ forest along these lines of sight is inconsistent with a uniform UVB model that is required to fit the observed lower envelope in \taueff\ at $z < 5.8$.  We emphasize that a simple rescaling of \fhi\ is unable to produce a reasonable fit to all the data at these redshifts.  For example, if we increase \fhi\ at $z \simeq 5.6$ by 0.45 dex verses the nominal value in equation~(\ref{eq:fHI}), the two-sided 95 per cent range in \taueff\ becomes $4.2 \le \tau_{\rm eff} \le 7.1$.  Although this would accommodate the \ulas0148\ value, the large majority of points near this redshift would then fall below the lower bound.

Before proceeding further, we note that although the \taueff\ values
at $z > 5.8$ are markedly higher than the data at $5.3 < z <
  5.8$, they do not on their own necessarily require an inhomogeneous
UVB.  The dotted lines in Fig.~\ref{fig:taueff_uniform} are for a
case where the neutral fraction evolves as $\langle f_{\rm H\,I} \rangle \propto
(1+z)^{15}$ at $z > 5.8$.  This is a somewhat arbitrary choice, but
the bounds in \taueff\ span the existing measurements and lower limits
at these redshifts.  Thus, while larger samples may ultimately require
a non-uniform UVB at $z > 5.8$, the present \lya\ data do not
currently demand it.  We note, however, that if the UVB
contains significant fluctuations over $5.3 < z < 5.8$, perhaps due to
variations in the mean free path (see below), then it is unlikely to
be uniform at higher redshifts.  Scatter in the UVB may also be
required to account for the range in \lyb\ opacities at $z \gtrsim 6$
measured by \citet{fan2006b}.

We can use our uniform UVB model to estimate the minimum amplitude of
UVB fluctuations required to explain the strongest outliers in
\taueff.  The dashed lines in Fig.~\ref{fig:taueff_uniform} show the
one-sided 95 per cent upper limit in \taueff\ expected when \fhi\ is
increased with respect to the nominal value (equation~\ref{eq:fHI}) by
0.15, 0.3, 0.45, and 0.6 dex (bottom to top).  The majority of points
can be accommodated by a factor of 2 increase in \fhi; however, the
two points for \ulas0148\ at $z = 5.63$ and 5.80 require an increase
in \fhi\ by a factor of $\gtrsim$3.  Note that the lower bound on \taueff\ will also increase when \fhi\ is increased, and so a higher \fhi\ will not accommodate the lowest-\taueff\ points.  In
Fig.~\ref{fig:taueff_uniform_trough} we replaced the
50~\hinvMpc\ points for \ulas0148\ with our lower limit for the
complete $\sim$110~\hinvMpc\ trough, while the dashed lines show the
expected upper limits for 100~\hinvMpc\ regions.  The complete trough
similarly requires a factor of $\gtrsim$3 increase in \fhi\ from our
nominal values.  Hence, it appears very likely that significant
fluctuations in \fhi\ in excess of those expected from density
  fluctuations alone must be present at $z \sim 5.6$--$5.8$.



\subsection{Galaxy UVB model}\label{sec:gal_uvb}

Some level of spatial variation in the UVB is expected simply due to the fact that the photons are emitted by discrete sources.  When the mean free path of ionizing photons is sufficiently long, however, the radiation field at any location will tend to reflect the contributions from a large number of sources, and so the amplitude of the fluctuations will be small unless there is a nearby, bright source such as a quasar.  On the other hand, recent measurements have shown that the mean free path decreases steeply with redshift over $2 \lesssim z \lesssim 6$ \citep{prochaska2009b,songaila2010,omeara2013,rudie2013,worseck2014}.  It is therefore possible that the fluctuations in the UVB at $z > 5$ we infer from the \lya\ forest are a natural consequence of the shortening of the mean free path.  

In this section we model the expected distribution in \taueff\ for
simple UVB models where star-forming galaxies are assumed to
provide the majority of ionizing photons at $z>5$ \citep[e.g.,][]{hm2012,robertson2013}.  AGN are not included, as their contribution to the UVB near 1 Ryd is believed to be small at these redshifts \citep[e.g.,][]{cowie2009}.  To construct the UVB we first
populate our simulation with sources and then calculate the intensity
of the radiation field as a function of position assuming a spatially
uniform mean free path.  Although this is clearly an
oversimplification, it gives us a first-order method for coupling the
radiation field to the density field, which is essential for
determining how fluctuations in the UVB affect the transmitted flux
statistics.  Below we assess whether this ``vanilla'' UVB model can
reproduce the observed variations in \taueff\ along different lines of
sight.

\begin{figure*}
   \centering
   \begin{minipage}{\textwidth}
   \begin{center}
   \includegraphics[width=0.6\textwidth]{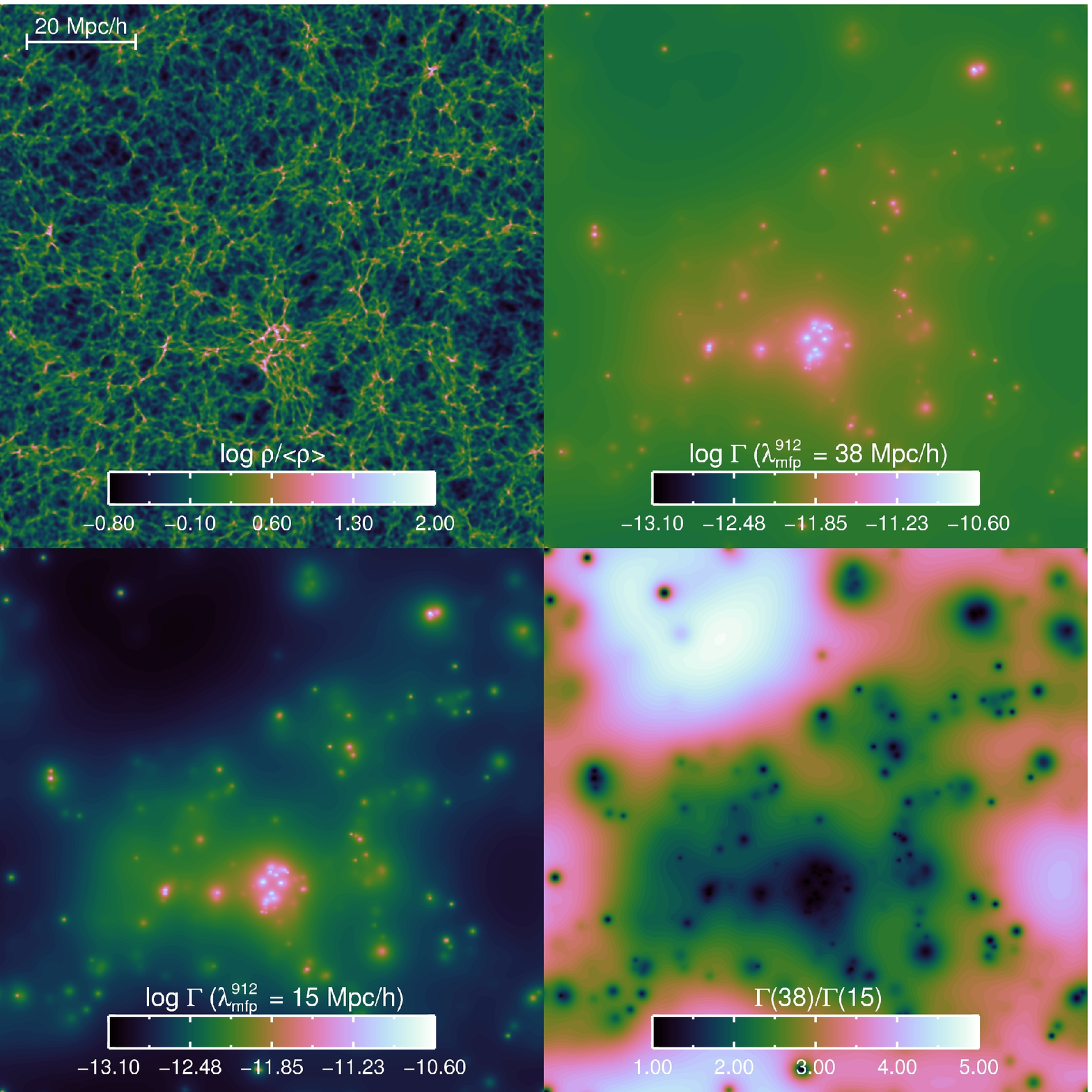}
   \vspace{-0.05in}
    \caption{A slice through our 100--1024 simulation at $z =
      5.62$.  The top left-hand panel shows the density field integrated
      over a 200~kpc$\,h^{-1}$ thick region. The top right-hand and bottom
      left-hand panels show the logarithm of the \hi\ ionization rate, where \G\ is in units of s$^{-1}$, for $\lambda_{\rm
        mfp}^{912} = 38$ and 13~\hinvMpc, respectively.  The bottom
      right-hand panel shows the ratio of the two ionization
      rates. Each panel is 100~\hinvMpc\ on a side.}
   \label{fig:uvb}
   \end{center}
   \end{minipage}
\end{figure*}

We model the UVB within our 100--1024 simulation following an approach
similar to that of \citet{boltonviel2011} and \citet{viel2013}.  First
we assign star-forming galaxies to dark matter haloes using an
abundance matching scheme \citep[e.g., equation 1 in][]{trenti2010}.
The haloes in the simulation volume were identified using a
  friends-of-friends halo finder, assuming a linking length of $0.2$
  and a minimum of 32 particles per halo.  We assume a galaxy duty
  cycle of unity in our calculations, but have verified that a lower
  values for the duty cycle of $0.5$ and $0.1$ have little effect on
  our results (see also Appendix~\ref{app:uvb_convergence}.)  

Parameters for the non-ionizing ($\lambda_{\rm rest} \sim 1500$~\AA)
UV luminosity function are determined by interpolating fits from
\citet[][]{bouwens2014}.  For our fiducial models we integrate down to
$M_{\rm AB} \le -18$, and assign luminosities to dark matter haloes by
randomly sampling the luminosity function.  Although this magnitude
limit neglects contributions from fainter sources, we find that the
impact on the \taueff\ distribution is converged at this limit (see
Appendix~\ref{app:uvb_convergence}).  We assume a galaxy spectral
energy distribution (SED) that is flat at $\lambda > 912$~\AA, follows a
power law $L_{\nu} \propto \nu^{-\alpha}$ at $\lambda < 912$~\AA, and
has a break at $\lambda = 912$~\AA\ of $A_{912} =
L_{\nu}(1500)/L_{\nu}(912)$.  For our fiducial model we adopt $\alpha
= 2$ and $A_{912} = 6.0$.  The amplitude of the UVB is then multiplied
by a scaling factor, \fion, which is chosen as described below.  This
factor nominally represents the escape fraction of ionizing photons;
however, in our models it is degenerate with $\alpha$ and $A_{912}$,
where for young stellar populations the latter may be a factor of 2
smaller than what we assume \citep[e.g.,][]{eldridge2012}.  It is also
degenerate with any contribution from galaxies fainter than $M_{\rm
  AB} = -18$.  Based on the luminosity functions of
\citet{bouwens2014}, these fainter galaxies may increase the total
emissivity by a factor of $\sim$2-3 over $4 < z < 6$.  Hence,
\fion\ may be up to a factor of $\sim$6 larger than the true
luminosity-weighted mean escape fraction at these redshifts.  For an
escape fraction $f_{\rm esc} \le 1$, therefore, $f_{\rm ion} \lesssim
6$ represents a reasonable upper limit for this parameter.  We note that it is obviously simplistic to assume that all galaxies have the same SED and escape fraction; if the escape fraction increases with luminosity, for example, we would expect larger fluctuations in the UVB.  We find, however, that even a model with contributions solely from galaxies with $M_{\rm AB} \le -21$ produces only a modest increase in the predicted scatter in \taueff\ (see Appendix~\ref{app:uvb_convergence}).

\begin{figure*}
   \centering
   \begin{minipage}{\textwidth}
   \begin{center}
   \includegraphics[width=0.7\textwidth]{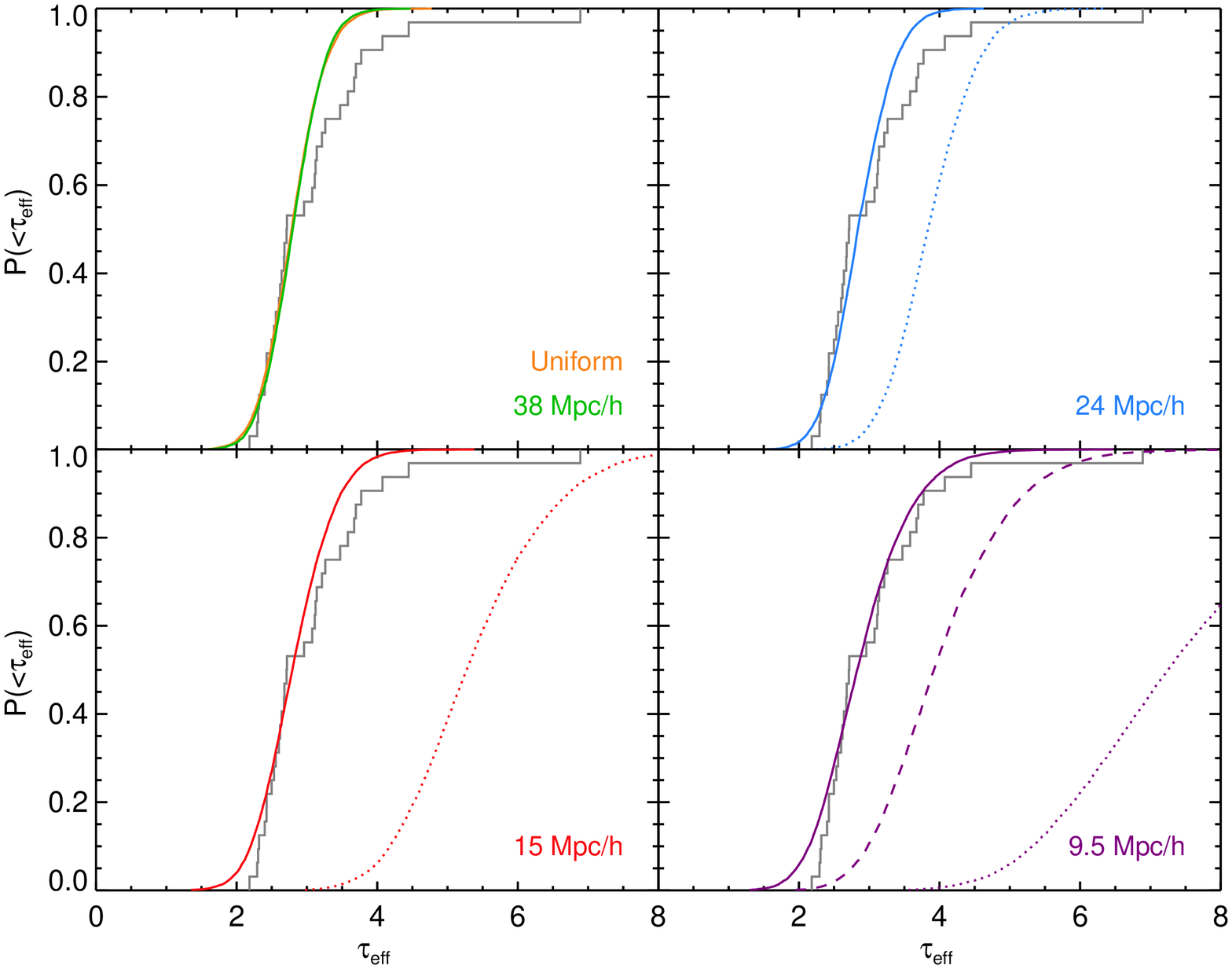}
   \vspace{-0.05in}
    \caption{Cumulative probability distribution function of
      \taueff\ near $z=5.6$.  The histogram in each panel shows
      \Ptaueff\ from the data over $5.5 < z < 5.7$.  Continuous lines
      give the predicted \Ptaueff\ from our galaxy UVB model for the
      mean free path indicated in the lower right-hand corner of each
      panel.  For the solid lines, \fion\ has been tuned individually
      for each \mfp\ such that \Ptaueff\ roughly matches the lower end
      of the data distribution ($f_{\rm ion} = 0.8$, 1.3, 2.3, and 4.0 for $\lambda_{\rm mfp}^{912} = 38$, 24, 15, and 9.5~\hinvMpc, respectively) .  The dotted lines in the top right-hand
      and bottom panels use $f_{\rm ion} = 0.8$, as in the $\lambda_{\rm mfp}^{912} = 38$~\hinvMpc\ case.  The dashed line in the lower right-hand panel is for $f_{\rm ion} = 2.3$.  In the top
      left-hand panel, \Ptaueff\ for a uniform UVB is also shown
      (orange line), although it is nearly indistinguishable from the
      galaxy UVB case (green line).}
   \label{fig:taueff_mfp}
   \end{center}
   \end{minipage}
\end{figure*}

For our fiducial mean free path evolution we adopt the fit from \citet{worseck2014}.  Using measurements over $2.4 \le z \le 5.2$ based on composite quasar spectra \citep{prochaska2009b,fumagalli2013,omeara2013,worseck2014} they find
\begin{equation}\label{eq:mfp}
   \lambda^{912}_{\rm mfp}(z) = 130 \left( \frac{1+z}{5} \right)^{-4.4}\,h^{-1}\,{\rm Mpc} \, ,
\end{equation}
where we have converted their values into comoving units.  Equation~(\ref{eq:mfp}) is broadly similar to, albeit somewhat steeper than, the evolution in \mfp\ found by \citet{songaila2010} out to $z \sim 6$.  These values are only taken as a reference point, however; a range of value in \mfp\ are explored below.  At each spatial position along our simulated lines of sight we then compute the specific intensity of the ionizing background between 1 and 4 Ryd by summing over the contribution from each galaxy as
\begin{equation}\label{eq:Jnu}
   J(\mathbf{r},\nu) = \frac{1}{4\pi} \sum\limits_{i=1}^N \frac{L_{i}(\mathbf{r}_i,\nu)}{4\pi | \mathbf{r}_i-\mathbf{r} |^2}
                                    e^{-\frac{| \mathbf{r}_i-\mathbf{r} |}{\lambda^{912}_{\rm mfp}} \left( \frac{\nu}{\nu_{912}}    
                                    \right)^{-3(\beta-1)}} \, .
\end{equation}
Here, $\nu_{912}$ is the frequency at the \hi\ ionizing edge, and $\beta$ is the slope of the \hi\ column density distribution, which sets the dependence of mean free path on frequency.  We adopt $\beta = 1.3$ \citep[e.g.,][]{songaila2010,becker2013b}.  The evolution in equation~(\ref{eq:mfp}) roughly agrees with that derived by \citet{songaila2010}.  In practice we perform the sum in equation~(\ref{eq:mfp}) for all sources with $| \mathbf{r}_i-\mathbf{r} | \le L_{\rm box}/2$ (within the periodic box), and add a contribution from larger distances assuming a spatially uniform distribution of sources.  Note that we  neglect the redshifting of ionizing photons, though this should have a relatively minor impact at $ z > 4$ \citep[e.g.,][]{becker2013b}.  The \hi\ photoionization rate is then computed as
\begin{equation}\label{eq:Gamma}
   \Gamma(\mathbf{r}) = 4\pi \int_{\nu_{912}}^{4\nu_{912}} \frac{d\nu}{h\nu} J(\mathbf{r},\nu) \sigma_{\rm H\, I}(\nu) \, ,
\end{equation}
where $\sigma_{\rm H\, I}(\nu)$ is the photoionization cross-section.

We focus our analysis on $z \simeq 5.6$, where the measured variation in \taueff\ is largest.  The density field and UVB at $z = 5.62$ for two values of \mfp\ are shown for a slice through our simulation box in Fig.~\ref{fig:uvb}.  As expected, the ionization rate correlates with the density, although for $\lambda_{\rm mfp}^{912} = 38$~\hinvMpc\ (the nominal value given by equation~\ref{eq:mfp}), the UVB in low density regions is still relatively uniform.  The mean intensity of the UVB scales as $\langle J \rangle \propto \lambda_{\rm mfp}^{912}$, but decreasing \mfp\ has the largest impact in low-density regions, which are least populated by ionizing sources.  This effect has the potential, at least, to increase the scatter in \taueff\ between lines of sight.

The predictions for our galaxy UVB model are compared to the data in
Fig.~\ref{fig:taueff_mfp}.  In each panel we plot the observed
cumulative probability distribution function, \Ptaueff, over $5.5 < z
< 5.7$.  Note that, for simplicity, we construct
  \Ptaueff\ treating lower limits as measurements, although we do not
  include the two lower limits from the \citet{fan2006b} data that
  fall below $\tau_{\rm eff} = 3$.  In cases where we have both lower
  and upper limits on \taueff\ we adopt their midpoint when
  constructing \Ptaueff.  We then overplot the expected \Ptaueff\ for
$\lambda_{\rm mfp}^{912} = 38$, 24, 15, and 9.5~\hinvMpc, which are
factors of 1.0, 0.63, 0.40 and 0.25 times the nominal value expected
from equation~(\ref{eq:mfp}).  The model distributions include a 20
per cent rms uncertainty in the continuum placement, meant to
mimic the effect of random continuum errors in the data.  The solid
lines show the model predictions when \fion\ is tuned such that
\Ptaueff\ roughly matches the lower end of the observed distribution.
The $\lambda_{\rm mfp}^{912} = 38$~\hinvMpc\ case uses $f_{\rm ion} =
0.8$, which is reasonable given the model parameters (see above).  The
$\lambda_{\rm mfp}^{912} = 9.5$~\hinvMpc\ case, however, uses $f_{\rm
  ion} = 4.0$, which is close to the expected upper limit  of $\sim$6 for
  this parameter.  Shorter mean free paths are therefore
probably not realistic for these models.  The $\lambda_{\rm mfp}^{912}
= 38$~\hinvMpc case produces nearly the same \Ptaueff\ as a uniform
UVB model, which is also plotted in the upper left-hand panel.  This
reflects the fact that the radiation field in the voids, which
dominate the transmission at $z > 5$, is relatively uniform for large
value of \mfp\ (e.g., Fig.~\ref{fig:uvb}).  Fluctuations in
  \taueff\ therefore remain dominated by variations in the density
  field \citep[see also][]{bolton2007,mesinger2009}. Both the
uniform UVB and $\lambda_{\rm mfp}^{912} = 38$~\hinvMpc\ models strongly
under-predict the number of high-\taueff\ lines of sight.  

The general agreement with observations does, in some sense, improve towards smaller values of \mfp\ (and correspondingly higher emissivities).  The model \Ptaueff\ for $\lambda_{\rm mfp}^{912} = 9.5$~\hinvMpc, $f_{\rm ion} = 4.0$  (lower right-hand panel, solid line) has the broadest distribution and roughly matches most of the data.  Even in this case, however, the probability of observing the highest \taueff\ value is essentially zero.  We also emphasize that this model requires an ionizing emissivity of $\sim$$5 \times 10^{51}~{\rm photons~s^{-1}~Mpc^{-3}}$, which is a factor of five higher than the most recent estimate at $z \simeq 4.8$ \citep{becker2013b}.  The dashed line in the lower right-hand panel is for $\lambda_{\rm mfp}^{912} = 9.5$~\hinvMpc, $f_{\rm ion} = 2.3$.  This is the only combination of parameters for which \Ptaueff\ is non-negligible for both the highest and lowest \taueff\ values in the data ($P(\le 2.2) = 0.005$, $P(\le 6.9) = 0.994$).  An Anderson-Darling test rejects the  hypothesis that the data were drawn from this distribution at $>$99.99 per cent confidence.  The remaining models are ruled out on the grounds that the predicted probabilities of observing the extreme values in the data are too small to be meaningfully calculated.


For the three cases with $\lambda_{\rm mfp}^{912} < 38$~\hinvMpc\
in Fig.~\ref{fig:taueff_mfp} we also show the predicted \Ptaueff\ when \fion\ is fixed to the value
used for the 38~\hinvMpc\ case (dotted lines).  As expected,
\Ptaueff\ shifts towards higher values of \taueff, yet a single
\mfp\ is again unable to match the full observed
\taueff\ distribution.  For this value of \fion, the lowest observed \taueff\ values only appear in the $\lambda_{\rm mfp}^{912} = 38$~\hinvMpc\ case, while the highest observed value is only predicted to occur with significant frequency when $\lambda_{\rm mfp}^{912} \le
15$~\hinvMpc.  Hence, for a given emissivity, fluctuations in
\mfp\ by factors of $\gtrsim$2.5 appear necessary to bracket the
observed \Ptaueff.




In summary, the failure of either a uniform UVB model or our simple
galaxy UVB model to reproduce the full distribution of
\taueff\ values, particularly near $z \simeq 5.6$, suggests that
more complicated ionization-driven fluctuations in the volume-averaged neutral
fraction are present at these redshifts.  Although variations in gas
temperature could technically produce variations in \lya\ opacity, the
high \taueff\ values towards \ulas0148\ would require those regions of
the IGM to be roughly a factor of 5 {\it colder} than average, a
scenario that is physically implausible in an ionized IGM.  We
therefore conclude that substantial ($\gtrsim$0.5 dex),
large-scale\footnote{Note, however, that it is difficult to
    quantify the exact scale on which fluctuations in the neutral
    fraction occur from the 1D line-of-sight data analysed here.
    Fluctuations which occur on smaller scales in 3D may appear to
    produce larger scale fluctuations in 1D due to aliasing
    \citep[e.g.,][]{mcquinn2011}.} (possibly $l \gtrsim 50$~\hinvMpc) fluctuations
in the neutral fraction must be present throughout at least part
of the IGM at these redshifts.  Given that the ionizing emissivity
from galaxies is likely to be comparatively uniform on these scales,
we expect that the \taueff\ fluctuations are driven primarily by
fluctuations in the mean free path.  We now turn towards examining the
\taueff\ data in more detail, and hence determining how these
fluctuations may be evolving with redshift.

\begin{figure*}
   \centering
   \begin{minipage}{\textwidth}
   \begin{center}
   \includegraphics[width=1.0\textwidth]{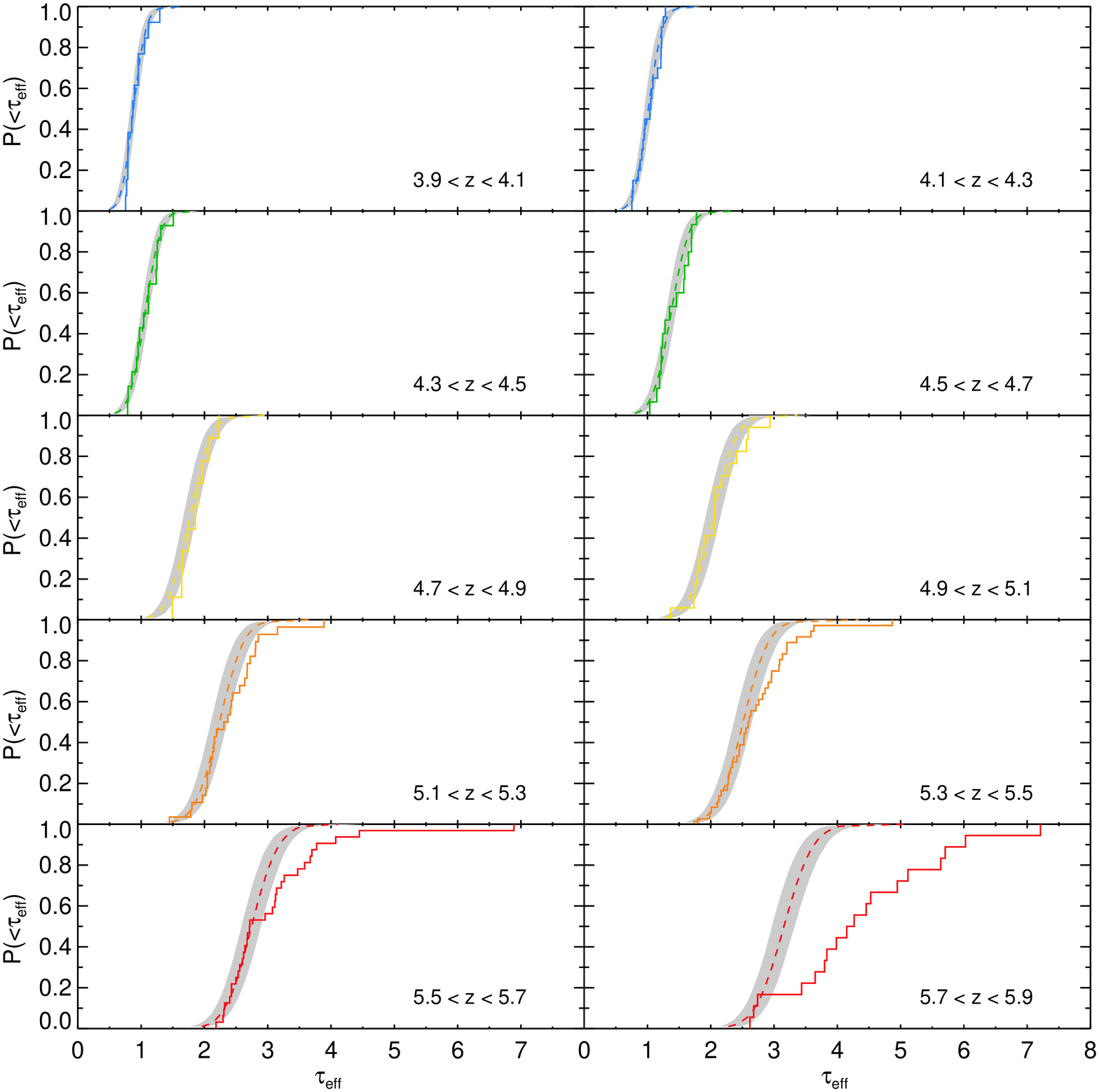}
   \vspace{-0.05in}
    \caption{Cumulative probability distribution function of
      \taueff\ in bins of $\Delta z = 0.2$.  The histograms in each
      panel are for the data.  Dashed lines give \Ptaueff\ for a
      uniform UVB (i.e., where the scatter in \taueff is driven by
        line-of-sight density fluctuations alone), where \fhi\ has
      been tuned such that the model \Ptaueff\ matches the lower end
      of the observed distribution.  The shaded regions show the
      effect of varying \fhi\ by $\pm$10 per cent.}
   \label{fig:taueff_distributions}
   \end{center}
   \end{minipage}
\end{figure*}

\section{Redshift Evolution of the \lya\ Opacity: Evidence for patchy Reionization}\label{sec:evolution}

In the previous sections we have argued that the
\lya\ \taueff\ distribution at $z \simeq 5.6$--$5.8$ is inconsistent
  with either line-of-sight density variations alone or a spatially
  fluctuating UVB with a fixed mean free path. We thus argue that
spatial variations in the mean free path must be
present at these redshifts.  As seen in
Fig.~\ref{fig:taueff_uniform}, however, the scatter in the observed
\taueff\ diminishes rapidly with redshift, until at $z \lesssim 5$ it
becomes consistent with that expected from fluctuations in the
  IGM density field alone.  We now investigate in more detail how the
\taueff\ distribution evolves with redshift.  As we demonstrate below,
simple models of the UVB, while they are unable to fully describe the
\taueff\ data at $z > 5$, can nevertheless provide insight into how
the IGM is evolving at these redshifts.

In Fig.~\ref{fig:taueff_distributions} we plot \Ptaueff\ for the
data over $3.9 < z < 5.9$ in redshift bins of $\Delta z = 0.2$.  For
each bin we then overplot \Ptaueff\ for a uniform UVB model with
\fhi\ tuned such that that the model matches the data over the maximum
possible range in \taueff\ starting at the low end.  By matching the low-\taueff\ end of the model to the low end of the data, these models represent the maximum \fhi\ that can reproduce the most transparent lines of sight.  As above, we can then investigate the extent to which these \fhi\ values also predict more opaque regions.  Note that
  this procedure differs from simply matching the global mean observed opacity
  with simulations, which implicitly assumes a uniform
  photo-ionization rate for gas probed by the {\it entire}
  \Ptaueff\ distribution, and thus ignores the additional scatter in
  the \taueff\ measurements and potentially underestimates the
  photo-ionization rate in the most highly ionized regions
  \citep[e.g.,][]{bolton2007,mesinger2009}.

For this comparison we use the 25--1024 simulation (i.e., two
simulated lines of sight are used per 50~\hinvMpc region), for which
we find \Ptaueff\ to be nearly converged with \fhi\ with respect
  to box size and mass resolution (see
Appendix~\ref{app:num_convergence}).  The model \taueff\ distributions
are interpolated between simulation output redshifts to match the
data.  The models also include an rms scatter in the continuum
with amplitude linearly interpolated between 10 and 20 percent between
$z=4$ and 5, and 20 per cent at $z > 5$.  The exact amplitude of the
continuum scatter is not critical to our analysis.  There may also be
systematic uncertainties in the continuum placement, however, which we
address below. Note again that we construct \Ptaueff\ treating lower limits as measurements.

Over $3.9 < z < 4.9$ the data are well matched by a uniform UVB model
over the full range in \taueff.  This agrees with the general
impression from Fig.~\ref{fig:taueff_uniform} that
  line-of-sight variations in the density field dominate the scatter
  in \taueff, which is perhaps not surprising
given the long mean free paths at these redshifts
(equation~\ref{eq:mfp}).  At $z > 4.9$ the data begin to diverge from
the uniform UVB model at the high-\taueff\ end.  We note, however,
that although the divergence increases with redshift, a substantial
fraction of the data remain consistent with the uniform UVB model up
to at least $z \sim 5.7$.  Over $5.5 < z < 5.7$, roughly half of the
data follow the expected \Ptaueff\ for a uniform UVB, even while the
remaining half follow an extended tail towards higher values.  Over
$5.7 < z < 5.9$, in contrast, less than 20 per cent of the data appear
to be consistent with density-driven fluctuations in \taueff.

The apparent agreement between much of data over $4.9 < z < 5.9$ and
the predicted \Ptaueff\ for a uniform UVB suggests that lines of sight
matched by the model may trace regions where the
  \hi\ photoionization rate is reasonably similar, at least in the
voids, which dominate the transmission at $z > 5$.  The fraction of
the \taueff\ data that require a somewhat lower
  photoionization rate, meanwhile, decreases rapidly with decreasing redshift
over this interval.  This trend is broadly consistent with the
final stages of patchy reionization
\citep[e.g.,][]{gnedin2000a,me2000,barkana2001}.  Even once the
ionized bubbles in a region of the IGM overlap and the volume-averaged
neutral fraction approaches zero, the local mean free path will still
evolve rapidly and exhibit a degree of spatial variance as
  residual patches of neutral hydrogen and/or Lyman limit systems at
  the edges of \hii\ regions are ionized
  \citep[e.g.,][]{furoh2005,choudhury2009a,alvarez2012,sobacchi2014}.  The
  final stages of reionization will progress until the local mean free
  path is set by large-scale structure rather than reionization
  topology.  At this point, fluctuations in Lyman limit opacity
  observed in the existing quasar spectra are primarily driven by variations in density rather
  than ionization, and the UVB in underdense regions will
  approach a global value that is relatively uniform.  We argue here that the fraction of the data in Fig.~\ref{fig:taueff_distributions} consistent with this natural end-point to reionization is small at $z \sim 6$ but approaches unity by $z \sim 5$.





\begin{figure}
   \begin{center}
   \includegraphics[width=0.45\textwidth]{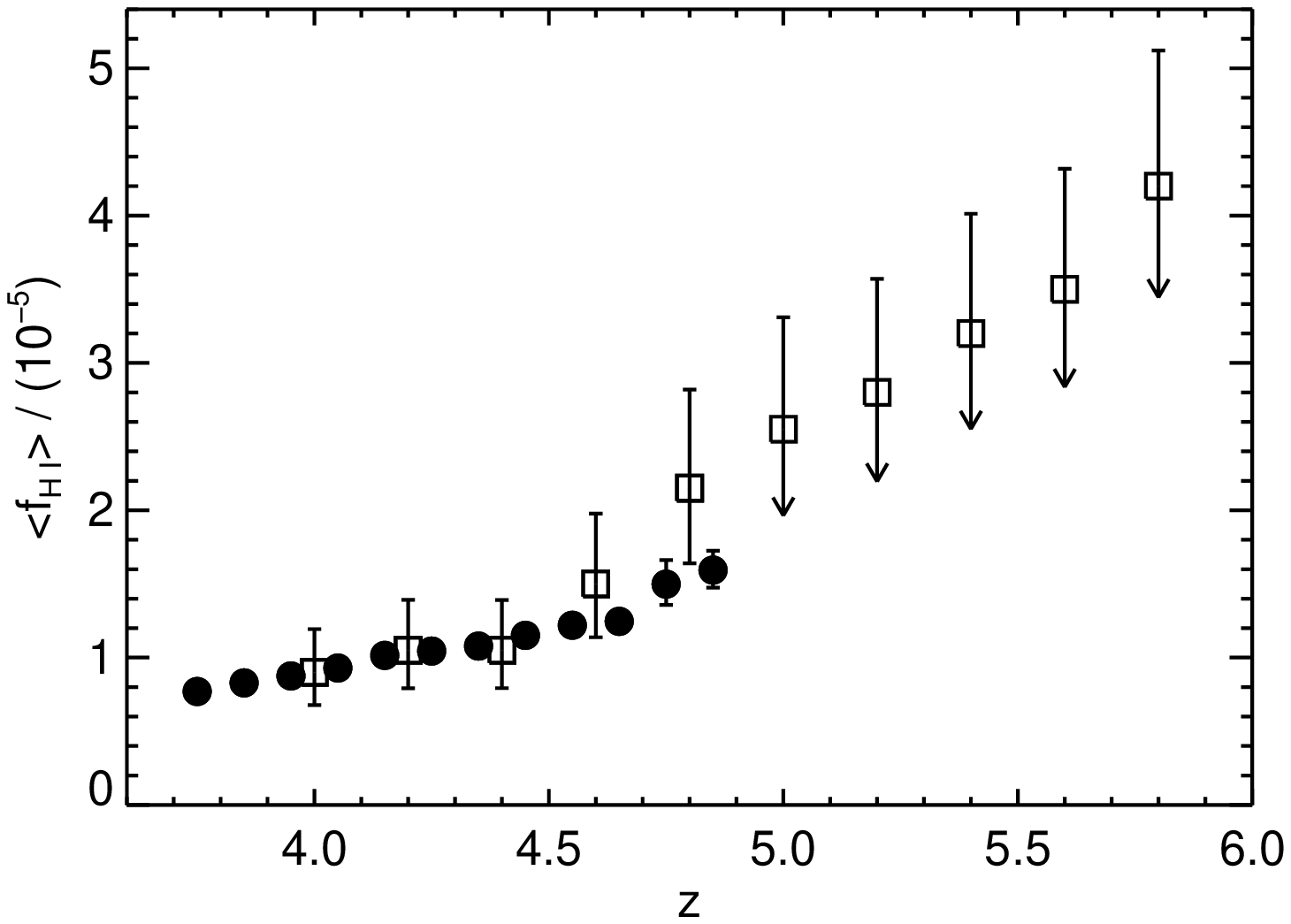}
   \vspace{-0.05in}
   \caption{Hydrogen neutral fraction in regions of the IGM where
       the line-of-sight variance in \taueff\ is well described by
       density fluctuations alone.  Open squares show values derived
     from the \Ptaueff\ fits in Fig.~\ref{fig:taueff_distributions}.
     Filled circles show values derived from the mean \lya\ opacity
     measurements of \citet{becker2013b}.}
   \label{fig:fHI}
   \end{center}
\end{figure}

The volume-weighted hydrogen
neutral fractions corresponding to the simulated
  \taueff\ distributions in Fig.~\ref{fig:taueff_distributions} are
shown in Fig.~\ref{fig:fHI}.  The error bars include possible
systematic errors in the quasar continua, which we take to be equal to
our adopted random continuum error estimates (5, 10 and 20 per cent at
$z=3$, 4 and 5, respectively, and 20 per cent at $z > 5$). These
  neutral fractions correspond to regions of the IGM where the
  line-of-sight variance in \taueff\ is consistent with density
  fluctuations alone.  At $z > 5$, since \fhi\ has been tuned to match only the low
end of the observed \taueff\ distribution, we are implicitly assuming that the matching regions
are generally of lower-than-average density.  If these regions are actually of higher density, then a higher ionization rate, and hence lower \fhi, would be required.  In this sense, the \fhi\ values at $z > 5$ in Fig.~\ref{fig:fHI} are upper limits.  We see,
nevertheless, that \fhi\ in these regions evolve gradually with
redshift, increasing by only a factor of 2 between $z \sim 5$ and 6.
This lends further support to the picture wherein lines of sight that
are consistent with the model \Ptaueff\ in
Fig.~\ref{fig:taueff_distributions} tend to probe regions of the IGM that have transitioned to a state 
where the mean free path is evolving relatively slowly.




\section{Summary}\label{sec:summary}

We have presented evidence for ionization-driven fluctuations in
the IGM neutral fraction near $z \sim 6$ based on an expanded set of
high-redshift quasar spectra.  The strongest evidence for fluctuations
is at $z \simeq 5.6 - 5.8$, where the deep \lya\ trough towards
\ulas0148\ contrasts strongly with the abundant transmitted flux
towards other lines of sight at the same redshift.  Using a suite of
large ($l_{\rm box} = 25$-100~\hinvMpc) hydrodynamical simulations, we
find that the full distribution of \taueff\ values cannot be
reproduced with either a uniform UVB or a simple background model that
assumes galaxies are the sources of ionizing photons and uses a fixed
mean free path.  

These data instead appear to require fluctuations in \fhi\ of at least
a factor of 3 on large scales.  These variations in \fhi\ must be
produced by fluctuations in the ionizing UV radiation field, which we
argue are likely to be driven by spatial variation in the local mean
free path throughout at least part of the IGM.  Our results are
broadly consistent with the original conclusions of \citet{fan2006b},
although we find that the Fan et al. data alone require only more
modest ($\lesssim$0.3 dex) fluctuations.  The new data presented here,
particularly the deep X-Shooter spectrum of \ulas0148, are essential
for motivating larger variations in \fhi.


The variations in \mfp\ argued for here are consistent with
expectations for the final stages of patchy hydrogen
reionization \citep{furoh2005,choudhury2009a,alvarez2012,sobacchi2014}.
During this transitional period the IGM can already be highly ionized
in a volume-averaged sense, yet the radiation field will be rapidly
evolving locally as residual Lyman limit systems and/or 
  remaining diffuse patches of neutral hydrogen are ionized.
  Based on the observed evolution of the \taueff\ distribution, we
  find that a decreasing fraction of the \taueff\ data towards higher redshift ($\lesssim$20
  per cent at $z \simeq 5.8$) is consistent
  with the variance expected from density fluctuations in the IGM
  alone.



Our analysis uses models of the radiation field that are purposefully
simplistic in order to allow us to assess whether fluctuations in the
UVB are required to explain the observed \taueff\ data, and how these
fluctuations may be evolving with redshift.  Predictions for
\Ptaueff\ from more sophisticated models include both sources and
sinks of ionizing photons in large volumes are clearly of interest for
developing a more nuanced picture of the IGM at these redshifts.  For
example, \citet{gnedin2014b} find a long tail towards high values of
\taueff\ at $z < 6$ in a set of simulations where \fhi\ approaches zero
between $z \simeq 6$ and 7.  These and
other simulations should help to translate the \taueff\ data
into more detailed constraints on reionization models.



\section*{Acknowledgements}

The authors thank Nick Gnedin, Martin Haehnelt, Fred Hamann, Paul Hewett, and Adam Lidz for helpful conversations, as well as Volker
  Springel for making {\sc gadget}-3 available.  This work is based in part on observations made with ESO Telescopes at the La Silla Paranal Observatory under program ID 084.A-0390.  Further observations were made at the W.M. Keck Observatory, which is operated as a scientific partnership between the California Institute of Technology and the University of California; it was made possible by the generous support of the W.M. Keck Foundation.  This paper also includes data gathered with the 6.5-m Magellan telescopes located at Las Campanas Observatory, Chile.
    The hydrodynamical simulations used in this work were performed
  using the Darwin Supercomputer of the University of Cambridge High
  Performance Computing Service (http://www.hpc.cam.ac.uk/), provided
  by Dell Inc. using Strategic Research Infrastructure Funding from
  the Higher Education Funding Council for England.  Fig.~\ref{fig:uvb} uses
  the cube helix colour scheme introduced by \cite{green2011}. GDB has been supported by an Ernest Rutherford Fellowship sponsored by the UK Science and Technology Facilities Council.  JSB
  acknowledges the support of a Royal Society University Research
  Fellowship.  PM acknowledges support by the NSF through grant OIA--1124453  and by NASA through grant NNX12AF87G.  BPV acknowledges funding through the ERC grant `Cosmic Dawn'.

\bibliographystyle{apj} \bibliography{uvb_fluctuations_refs}

\begin{thebibliography}{73}
\expandafter\ifx\csname natexlab\endcsname\relax\def\natexlab#1{#1}\fi

\bibitem[{Allen {et~al.}(2011)Allen, Hewett, Maddox, Richards, \&
  Belokurov}]{allen2011}
Allen, J.~T., Hewett, P.~C., Maddox, N., Richards, G.~T., \& Belokurov, V.
  2011, \mnras, 410, 860

\bibitem[{Alvarez \& Abel(2012)}]{alvarez2012}
Alvarez, M.~A., \& Abel, T. 2012, \apj, 747, 126

\bibitem[{Ba{\~n}ados {et~al.}(2014)Ba{\~n}ados, Venemans, Morganson, Decarli,
  Walter, Chambers, Rix, Farina, Fan, Jiang, McGreer, Rosa, Simcoe, Wei{\ss},
  Price, Morgan, Burgett, Greiner, Kaiser, Kudritzki, Magnier, Metcalfe,
  Stubbs, Sweeney, Tonry, Wainscoat, \& Waters}]{banados2014}
Ba{\~n}ados, E., {et~al.} 2014, \aj, 148, 14

\bibitem[{Barkana \& Loeb(2001)}]{barkana2001}
Barkana, R., \& Loeb, A. 2001, Phys. Rep., 349, 125

\bibitem[{Baskin {et~al.}(2013)Baskin, Laor, \& Hamann}]{baskin2013}
Baskin, A., Laor, A., \& Hamann, F. 2013, \mnras, 432, 1525

\bibitem[{Becker \& Bolton(2013)}]{becker2013b}
Becker, G.~D., \& Bolton, J.~S. 2013, \mnras, 436, 1023

\bibitem[{Becker {et~al.}(2011{\natexlab{a}})Becker, Bolton, Haehnelt, \&
  Sargent}]{becker2011a}
Becker, G.~D., Bolton, J.~S., Haehnelt, M.~G., \& Sargent, W. L.~W.
  2011{\natexlab{a}}, \mnras, 410, 1096

\bibitem[{Becker {et~al.}(2013)Becker, Hewett, Worseck, \&
  Prochaska}]{becker2013a}
Becker, G.~D., Hewett, P.~C., Worseck, G., \& Prochaska, J.~X. 2013, \mnras,
  430, 2067

\bibitem[{Becker {et~al.}(2009)Becker, Rauch, \& Sargent}]{becker2009}
Becker, G.~D., Rauch, M., \& Sargent, W. L.~W. 2009, \apj, 698, 1010

\bibitem[{Becker {et~al.}(2011{\natexlab{b}})Becker, Sargent, Rauch, \&
  Calverley}]{becker2011b}
Becker, G.~D., Sargent, W. L.~W., Rauch, M., \& Calverley, A.~P.
  2011{\natexlab{b}}, \apj, 735, 93

\bibitem[{Becker {et~al.}(2012)Becker, Sargent, Rauch, \&
  Carswell}]{becker2012}
Becker, G.~D., Sargent, W. L.~W., Rauch, M., \& Carswell, R.~F. 2012, \apj,
  744, 91

\bibitem[{Becker {et~al.}(2006)Becker, Sargent, Rauch, \& Simcoe}]{becker2006}
Becker, G.~D., Sargent, W. L.~W., Rauch, M., \& Simcoe, R.~A. 2006, \apj, 640,
  69

\bibitem[{Becker {et~al.}(2001)Becker, Fan, White, Strauss, Narayanan, Lupton,
  Gunn, Annis, Bahcall, Brinkmann, Connolly, Csabai, Czarapata, Doi, Heckman,
  Hennessy, Ivezi{\'c}, Knapp, Lamb, McKay, Munn, Nash, Nichol, Pier, Richards,
  Schneider, Stoughton, Szalay, Thakar, \& York}]{becker2001}
Becker, R.~H., {et~al.} 2001, \aj, 122, 2850

\bibitem[{Bolton \& Becker(2009)}]{boltonbecker2009}
Bolton, J.~S., \& Becker, G.~D. 2009, \mnras, 398, L26

\bibitem[{Bolton \& Haehnelt(2007)}]{bolton2007}
Bolton, J.~S., \& Haehnelt, M.~G. 2007, \mnras, 382, 325

\bibitem[{Bolton {et~al.}(2005)Bolton, Haehnelt, Viel, \&
  Springel}]{bolton2005}
Bolton, J.~S., Haehnelt, M.~G., Viel, M., \& Springel, V. 2005, \mnras, 357,
  1178

\bibitem[{Bolton \& Viel(2011)}]{boltonviel2011}
Bolton, J.~S., \& Viel, M. 2011, \mnras, 414, 241

\bibitem[{Bouwens {et~al.}(2007)Bouwens, Illingworth, Franx, \&
  Ford}]{bouwens2007}
Bouwens, R.~J., Illingworth, G.~D., Franx, M., \& Ford, H. 2007, \apj, 670, 928

\bibitem[{Bouwens {et~al.}(2014)Bouwens, Illingworth, Oesch, Trenti, Labbe',
  Bradley, Carollo, van Dokkum, Gonzalez, Holwerda, Franx, Spitler, Smit, \&
  Magee}]{bouwens2014}
Bouwens, R.~J., {et~al.} 2014, arXiv:1403.4295

\bibitem[{Calverley {et~al.}(2011)Calverley, Becker, Haehnelt, \&
  Bolton}]{calverley2011}
Calverley, A.~P., Becker, G.~D., Haehnelt, M.~G., \& Bolton, J.~S. 2011,
  \mnras, 412, 2543

\bibitem[{Chornock {et~al.}(2014)Chornock, Berger, Fox, Fong, Laskar, \&
  Roth}]{chornock2014}
Chornock, R., Berger, E., Fox, D.~B., Fong, W., Laskar, T., \& Roth, K.~C.
  2014, arXiv:1405.7400

\bibitem[{Chornock {et~al.}(2013)Chornock, Berger, Fox, Lunnan, Drout, fai
  Fong, Laskar, \& Roth}]{chornock2013}
Chornock, R., Berger, E., Fox, D.~B., Lunnan, R., Drout, M.~R., fai Fong, W.,
  Laskar, T., \& Roth, K.~C. 2013, \apj, 774, 26

\bibitem[{Choudhury {et~al.}(2009)Choudhury, Haehnelt, \&
  Regan}]{choudhury2009a}
Choudhury, T.~R., Haehnelt, M.~G., \& Regan, J. 2009, \mnras, 394, 960

\bibitem[{Cooke {et~al.}(2011)Cooke, Pettini, Steidel, Rudie, \&
  Nissen}]{cooke2011b}
Cooke, R., Pettini, M., Steidel, C.~C., Rudie, G.~C., \& Nissen, P.~E. 2011,
  \mnras, 417, 1534

\bibitem[{Cowie {et~al.}(2009)Cowie, Barger, \& Trouille}]{cowie2009}
Cowie, L.~L., Barger, A.~J., \& Trouille, L. 2009, \apj, 692, 1476

\bibitem[{Djorgovski {et~al.}(2001)Djorgovski, Castro, Stern, \&
  Mahabal}]{djorgovski2001}
Djorgovski, S.~G., Castro, S., Stern, D., \& Mahabal, A.~A. 2001, \apj, 560, L5

\bibitem[{D'Odorico {et~al.}(2006)D'Odorico, Dekker, Mazzoleni, Vernet,
  Guinouard, Groot, Hammer, Rasmussen, Kaper, Navarro, Pallavicini, Peroux, \&
  Zerbi}]{sdodorico2006}
D'Odorico, S., {et~al.} 2006, \procspie, 6269, 626933

\bibitem[{D'Odorico {et~al.}(2013)D'Odorico, Cupani, Cristiani, Maiolino,
  Molaro, Nonino, Centuri{\'o}n, Cimatti, di~Serego~Alighieri, Fiore, Fontana,
  Gallerani, Giallongo, Mannucci, Marconi, Pentericci, Viel, \&
  Vladilo}]{dodorico2013}
D'Odorico, V., {et~al.} 2013, \mnras, 435, 1198

\bibitem[{Eisenstein \& Hu(1999)}]{eisensteinhu1999}
Eisenstein, D.~J., \& Hu, W. 1999, \apj, 511, 5

\bibitem[{Eldridge \& Stanway(2012)}]{eldridge2012}
Eldridge, J.~J., \& Stanway, E.~R. 2012, \mnras, 419, 479

\bibitem[{Fan {et~al.}(2002)Fan, Narayanan, Strauss, White, Becker, Pentericci,
  \& Rix}]{fan2002}
Fan, X., Narayanan, V.~K., Strauss, M.~A., White, R.~L., Becker, R.~H.,
  Pentericci, L., \& Rix, H.-W. 2002, \aj, 123, 1247

\bibitem[{Fan {et~al.}(2006)Fan, Strauss, Becker, White, Gunn, Knapp, Richards,
  Schneider, Brinkmann, \& Fukugita}]{fan2006b}
Fan, X., {et~al.} 2006, \aj, 132, 117

\bibitem[{Faucher-Gigu{\`e}re {et~al.}(2008)Faucher-Gigu{\`e}re, Lidz,
  Hernquist, \& Zaldarriaga}]{fg2008b}
Faucher-Gigu{\`e}re, C.-A., Lidz, A., Hernquist, L., \& Zaldarriaga, M. 2008,
  \apj, 682, L9

\bibitem[{Fumagalli {et~al.}(2013)Fumagalli, O'Meara, Prochaska, \&
  Worseck}]{fumagalli2013}
Fumagalli, M., O'Meara, J.~M., Prochaska, J.~X., \& Worseck, G. 2013, \apj,
  775, 78

\bibitem[{Furlanetto \& Oh(2005)}]{furoh2005}
Furlanetto, S.~R., \& Oh, S.~P. 2005, \mnras, 363, 1031

\bibitem[{Gnedin(2000)}]{gnedin2000a}
Gnedin, N.~Y. 2000, \apj, 535, 530

\bibitem[{Gnedin \& Kaurov(2014)}]{gnedin2014b}
Gnedin, N.~Y., \& Kaurov, A.~A. 2014, \apj, 793, 30

\bibitem[{Green(2011)}]{green2011}
Green, D.~A. 2011, Bull. Astron. Soc. India, 39, 289

\bibitem[{Haardt \& Madau(2001)}]{hm2001}
Haardt, F., \& Madau, P. 2001, in Clusters of Galax- ies and the High Redshift
  Universe Observed in X- rays, Neumann, D. M. {\&} Tran, J. T. V. ed.,
  arXiv:0106018

\bibitem[{Haardt \& Madau(2012)}]{hm2012}
---. 2012, \apj, 746, 125

\bibitem[{Horne(1986)}]{horne1986}
Horne, K. 1986, \pasp, 98, 609

\bibitem[{Keating {et~al.}(2014)Keating, Haehnelt, Becker, \&
  Bolton}]{keating2014}
Keating, L.~C., Haehnelt, M.~G., Becker, G.~D., \& Bolton, J.~S. 2014, \mnras,
  438, 1820

\bibitem[{Kelson(2003)}]{kelson2003}
Kelson, D.~D. 2003, \pasp, 115, 688

\bibitem[{Kim {et~al.}(2007)Kim, Bolton, Viel, Haehnelt, \& Carswell}]{kim2007}
Kim, T.-S., Bolton, J.~S., Viel, M., Haehnelt, M.~G., \& Carswell, R.~F. 2007,
  \mnras, 382, 1657

\bibitem[{Kirkman {et~al.}(2005)Kirkman, Tytler, Suzuki, Melis, Hollywood,
  James, So, Lubin, Jena, Norman, \& Paschos}]{kirkman2005}
Kirkman, D., {et~al.} 2005, \mnras, 360, 1373

\bibitem[{Lawrence {et~al.}(2007)Lawrence, Warren, Almaini, Edge, Hambly,
  Jameson, Lucas, Casali, Adamson, Dye, Emerson, Foucaud, Hewett, Hirst,
  Hodgkin, Irwin, Lodieu, McMahon, Simpson, Smail, Mortlock, \&
  Folger}]{lawrence2007}
Lawrence, A., {et~al.} 2007, \mnras, 379, 1599

\bibitem[{Lidz {et~al.}(2007)Lidz, McQuinn, Zaldarriaga, Hernquist, \&
  Dutta}]{lidz2007}
Lidz, A., McQuinn, M., Zaldarriaga, M., Hernquist, L., \& Dutta, S. 2007, \apj,
  670, 39

\bibitem[{Lidz {et~al.}(2006)Lidz, Oh, \& Furlanetto}]{lidz2006}
Lidz, A., Oh, S.~P., \& Furlanetto, S.~R. 2006, \apj, 639, L47

\bibitem[{Lu {et~al.}(1996)Lu, Sargent, Womble, \& Barlow}]{lu1996}
Lu, L., Sargent, W. L.~W., Womble, D.~S., \& Barlow, T.~A. 1996, \apjl, 457, L1

\bibitem[{McQuinn {et~al.}(2011)McQuinn, Hernquist, Lidz, \&
  Zaldarriaga}]{mcquinn2011}
McQuinn, M., Hernquist, L., Lidz, A., \& Zaldarriaga, M. 2011, \mnras, 415, 977

\bibitem[{Mesinger(2010)}]{mesinger2010}
Mesinger, A. 2010, \mnras, 407, 1328

\bibitem[{Mesinger \& Furlanetto(2009)}]{mesinger2009}
Mesinger, A., \& Furlanetto, S. 2009, \mnras, 400, 1461

\bibitem[{Miralda-Escud{\'e} {et~al.}(2000)Miralda-Escud{\'e}, Haehnelt, \&
  Rees}]{me2000}
Miralda-Escud{\'e}, J., Haehnelt, M., \& Rees, M.~J. 2000, \apj, 530, 1

\bibitem[{Mortlock {et~al.}(2009)Mortlock, Patel, Warren, Venemans, McMahon,
  Hewett, Simpson, Sharp, Burningham, Dye, Ellis, Gonzales-Solares, \&
  Hu{\'e}lamo}]{mortlock2009}
Mortlock, D.~J., {et~al.} 2009, \aap, 505, 97

\bibitem[{O'Meara {et~al.}(2013)O'Meara, Prochaska, Worseck, Chen, \&
  Madau}]{omeara2013}
O'Meara, J.~M., Prochaska, J.~X., Worseck, G., Chen, H.-W., \& Madau, P. 2013,
  \apj, 765, 137

\bibitem[Planck Collaboration XVI(2014)]{planckXVI} Planck Collaboration XVI 2014, \aap, 571, A16

\bibitem[{Prochaska {et~al.}(2009)Prochaska, Worseck, \&
  O'Meara}]{prochaska2009b}
Prochaska, J.~X., Worseck, G., \& O'Meara, J.~M. 2009, \apjl, 705, L113

\bibitem[{Robertson {et~al.}(2013)Robertson, Furlanetto, Schneider, Charlot,
  Ellis, Stark, McLure, Dunlop, Koekemoer, Schenker, Ouchi, Ono, Curtis-Lake,
  Rogers, Bowler, \& Cirasuolo}]{robertson2013}
Robertson, B.~E., {et~al.} 2013, \apj, 768, 71

\bibitem[{Rudie {et~al.}(2013)Rudie, Steidel, Shapley, \& Pettini}]{rudie2013}
Rudie, G.~C., Steidel, C.~C., Shapley, A.~E., \& Pettini, M. 2013, \apj, 769,
  146

\bibitem[{Ryan-Weber {et~al.}(2009)Ryan-Weber, Pettini, Madau, \&
  Zych}]{rw2009}
Ryan-Weber, E.~V., Pettini, M., Madau, P., \& Zych, B.~J. 2009, \mnras, 395,
  1476

\bibitem[{Schaye {et~al.}(2003)Schaye, Aguirre, Kim, Theuns, Rauch, \&
  Sargent}]{schaye2003}
Schaye, J., Aguirre, A., Kim, T.-S., Theuns, T., Rauch, M., \& Sargent, W.
  L.~W. 2003, \apj, 596, 768

\bibitem[{Simcoe {et~al.}(2011)Simcoe, Cooksey, Matejek, Burgasser, Bochanski,
  Lovegrove, Bernstein, Pipher, Forrest, McMurtry, Fan, \&
  O'Meara}]{simcoe2011a}
Simcoe, R.~A., {et~al.} 2011, \apj, 743, 21

\bibitem[{Sobacchi \& Mesinger(2014)}]{sobacchi2014}
Sobacchi, E., \& Mesinger, A. 2014, \mnras, 440, 1662

\bibitem[{Songaila(2004)}]{songaila2004}
Songaila, A. 2004, \aj, 127, 2598

\bibitem[{Songaila \& Cowie(2010)}]{songaila2010}
Songaila, A., \& Cowie, L.~L. 2010, \apj, 721, 1448

\bibitem[{Springel(2005)}]{springel2005}
Springel, V. 2005, \mnras, 364, 1105

\bibitem[{Theuns {et~al.}(1998)Theuns, Leonard, Efstathiou, Pearce, \&
  Thomas}]{theuns1998}
Theuns, T., Leonard, A., Efstathiou, G., Pearce, F.~R., \& Thomas, P.~A. 1998,
  \mnras, 301, 478

\bibitem[{Trenti {et~al.}(2010)Trenti, Stiavelli, Bouwens, Oesch, Shull,
  Illingworth, Bradley, \& Carollo}]{trenti2010}
Trenti, M., Stiavelli, M., Bouwens, R.~J., Oesch, P., Shull, J.~M.,
  Illingworth, G.~D., Bradley, L.~D., \& Carollo, C.~M. 2010, \apjl, 714, L202

\bibitem[{Viel {et~al.}(2013)Viel, Becker, Bolton, \& Haehnelt}]{viel2013}
Viel, M., Becker, G.~D., Bolton, J.~S., \& Haehnelt, M.~G. 2013, \prd, 88,
  43502

\bibitem[{Wang {et~al.}(2013)Wang, Wagg, Carilli, Walter, Lentati, Fan,
  Riechers, Bertoldi, Narayanan, Strauss, Cox, Omont, Menten, Knudsen, Neri, \&
  Jiang}]{wang2013}
Wang, R., {et~al.} 2013, \apj, 773, 44

\bibitem[{White {et~al.}(2003)White, Becker, Fan, \& Strauss}]{white2003}
White, R.~L., Becker, R.~H., Fan, X., \& Strauss, M.~A. 2003, \aj, 126, 1

\bibitem[{Worseck {et~al.}(2014)Worseck, Prochaska, O'Meara, Becker, Ellison,
  Lopez, Meiksin, M{\'e}nard, Murphy, \& Fumagalli}]{worseck2014}
Worseck, G., {et~al.} 2014, \mnras, 445, 1745

\bibitem[{Wyithe \& Bolton(2011)}]{wyithe2011}
Wyithe, J. S.~B., \& Bolton, J.~S. 2011, \mnras, 412, 1926

\end{thebibliography}

\appendix

\section{Convergence Tests}\label{app:convergence}

\begin{figure*}
   \centering
   \begin{minipage}{\textwidth}
   \begin{center}
   \includegraphics[width=0.8\textwidth]{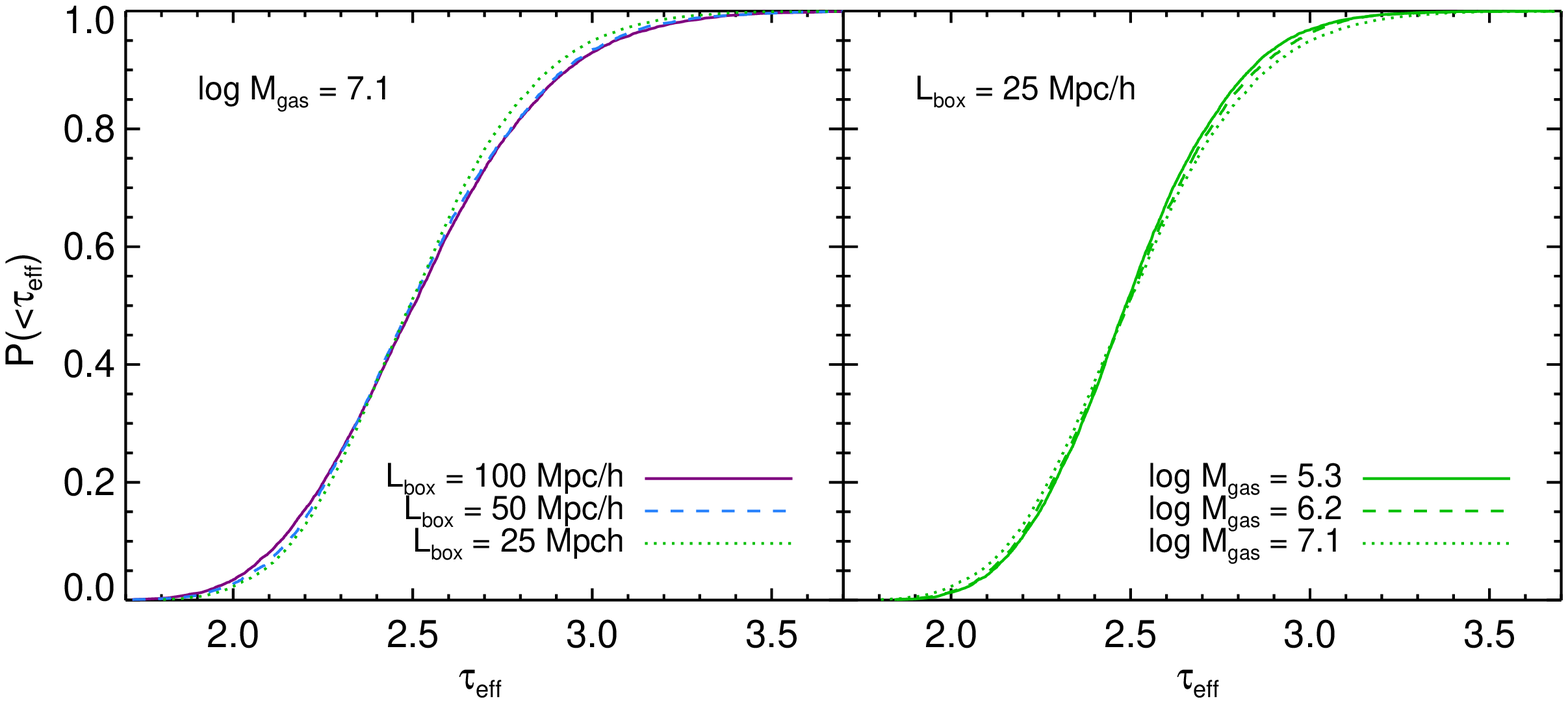}
   \vspace{-0.05in}
    \caption{Numerical convergence of \Ptaueff\ for fixed mean
      \lya\ flux.  For each simulation we compute \taueff\ values at
      $z=5.62$ over 50~\hinvMpc\ lines of sight after rescaling the
      photoionization rate such that $\langle F \rangle = 0.084$.  The
      left-hand panel shows the effects of varying the simulation
      box size while fixing the gas particle mass to $M_{\rm gas} =
      1.1 \times 10^7~{\rm M_{\odot}}\,h^{-1}$, while the right-hand
      panel shows the effects of varying the mass resolution for a
      fixed box size, $l_{\rm box} = 25$~\hinvMpc.}
   \label{fig:fixed_fmean}
   \end{center}
   \end{minipage}
\end{figure*}

\begin{figure*}
   \centering
   \begin{minipage}{\textwidth}
   \begin{center}
   \includegraphics[width=0.65\textwidth]{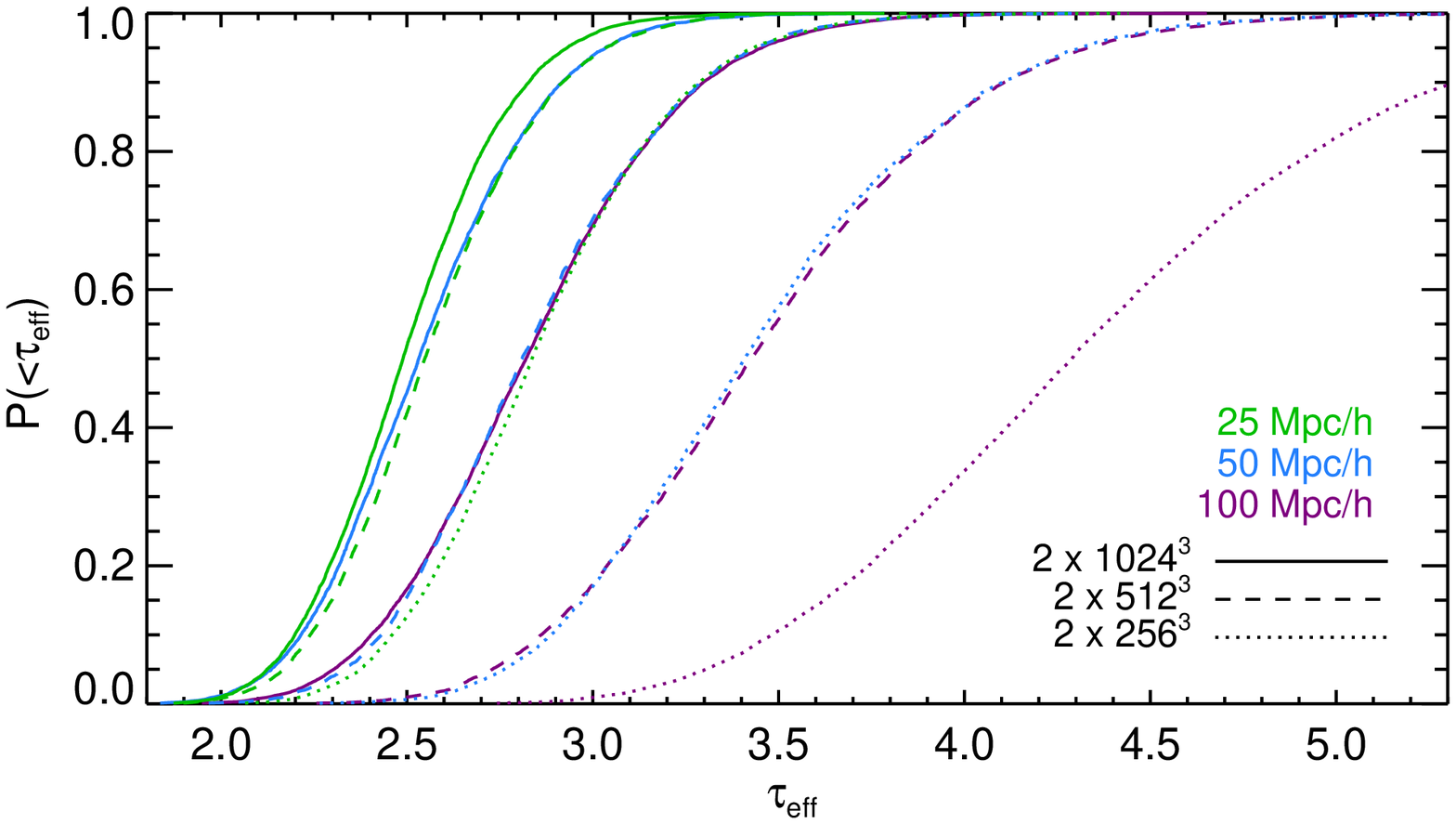}
   \vspace{-0.05in}
    \caption{Numerical convergence of \Ptaueff\ for fixed hydrogen neutral fraction.  For each simulation we compute \taueff\ values at $z=5.62$ over 50~\hinvMpc\ lines of sight after rescaling the photoionization rate such that $\langle f_{\rm H\,I} \rangle = 2.9 \times 10^{-5}$.  Green, blue, and purple lines are for $\l_{\rm box} = 25$, 50, and 100~\hinvMpc, respectively, while solid, dashed, and dotted lines are for total (dark matter plus gas) particle numbers $n = 2 \times 1024^3$, $2 \times 512^3$, and $2 \times 256^3$, respectively.  Runs with similar mass resolution produce similar \Ptaueff\ results, with only a weak dependence on box size.  Decreasing the mass resolution, however, can produce a significantly more opaque \lya\ forest at this redshift.}
   \label{fig:fixed_fHI}
   \end{center}
   \end{minipage}
\end{figure*}

In this appendix we address several issues related to numerical
convergence.  As discussed in \citet{boltonbecker2009}, the
\lya\ forest becomes increasingly sensitive to box size and mass
resolution towards higher redshifts, since the transmission becomes
dominated by rare voids.  We therefore focus our tests at $z \simeq
5.6$, which is both near the upper end of the redshift range probed in
this paper and the redshift where the largest range in \taueff\ values
are observed.  For our convergence tests we use a suite of nine
simulations with box sizes that span 25--100~\hinvMpc, and gas
particle masses in the range $1.79 \times 10^5$--$7.34 \times
10^8~M_\odot\,h^{-1}$.  These are listed in Table~\ref{tab:hydrosims}.
Except where noted we compute \taueff\ over 50~\hinvMpc\ lines of
sight, hence for the 25~\hinvMpc\ boxes we join two randomly chosen
lines of sight per measurement, whereas for the 100~\hinvMpc\ box we
extract two measurements per line of sight.

\subsection{Numerical convergence}\label{app:num_convergence}

We begin by examining the convergence of our simulated \Ptaueff\ with
box size and mass resolution for a uniform UVB.  In
Fig.~\ref{fig:fixed_fmean} we plot \Ptaueff\ at $z = 5.62$ where the
mean transmitted \lya\ flux is fixed to $\langle F \rangle = 0.084$
for all simulations.  At fixed $\langle F \rangle$ the simulated
\Ptaueff\ increases marginally with box size (left-hand panel), though
there is little difference between the 50 and 100~\hinvMpc\ boxes.
\Ptaueff\ is somewhat narrower for smaller gas particles masses
(right-hand panel), consistent with expectations from
\citet{boltonbecker2009}.  Our choice of a 100~\hinvMpc\ box with
$M_{\rm gas} = 1.15 \times 10^7~M_\odot\,h^{-1}$ is therefore
conservative in terms of determining whether the observed scatter in
\taueff\ can be reproduced using a uniform UVB.

\begin{figure}
   \begin{center}
   \includegraphics[width=0.45\textwidth]{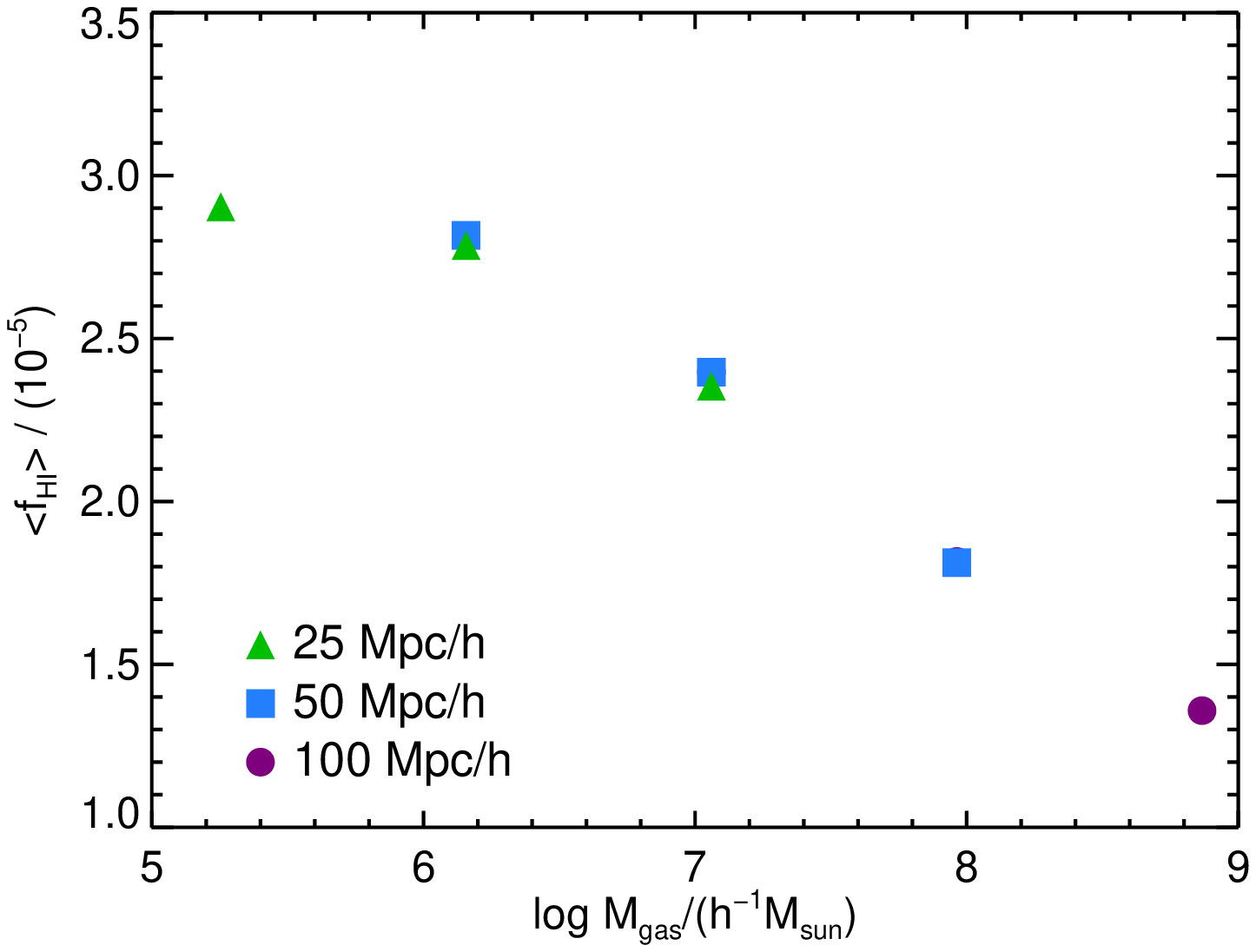}
   \vspace{-0.05in}
   \caption{Convergence of the volume-averaged hydrogen neutral fraction with box size and mass resolution.  For each simulation, \fhi\ at $z=5.62$ is computed after rescaling the photoionization rate to produce a fixed mean \lya\ flux, $\langle F \rangle = 0.084$.}
   \label{fig:neutralfrac_convergence}
   \end{center}
\end{figure}

The results are somewhat different if we evaluate \Ptaueff\ at a fixed hydrogen neutral fraction.  We plot \Ptaueff\ at $z = 5.62$ for our nine numerical convergence runs in Fig.~\ref{fig:fixed_fHI}, where for each run we have fixed $\langle f_{\rm H\,I} \rangle = 2.9 \times 10^{-5}$.  Although the predicted \taueff\ distribution shows relatively little dependence on box size, it is strongly sensitive to mass resolution.  Runs using smaller gas particle masses generate voids that are more transparent  \citep[see discussion in][]{boltonbecker2009}.  In Fig.~\ref{fig:neutralfrac_convergence} we plot \fhi\ at a fixed $\langle F \rangle = 0.084$.  This again shows relatively little dependence on box size over 25-100~\hinvMpc, but a strong dependence on mass resolution.  The neutral fraction appears to be roughly converged for our 25--1024 run ($M_{\rm gas} = 1.8 \times 10^5~M_\odot\,h^{-1}$), which we used to measure the \fhi\ values shown in Fig.~\ref{fig:fHI}.

\subsection{Cosmology}\label{app:cosmology}

\begin{figure}
   \begin{center}
   \includegraphics[width=0.45\textwidth]{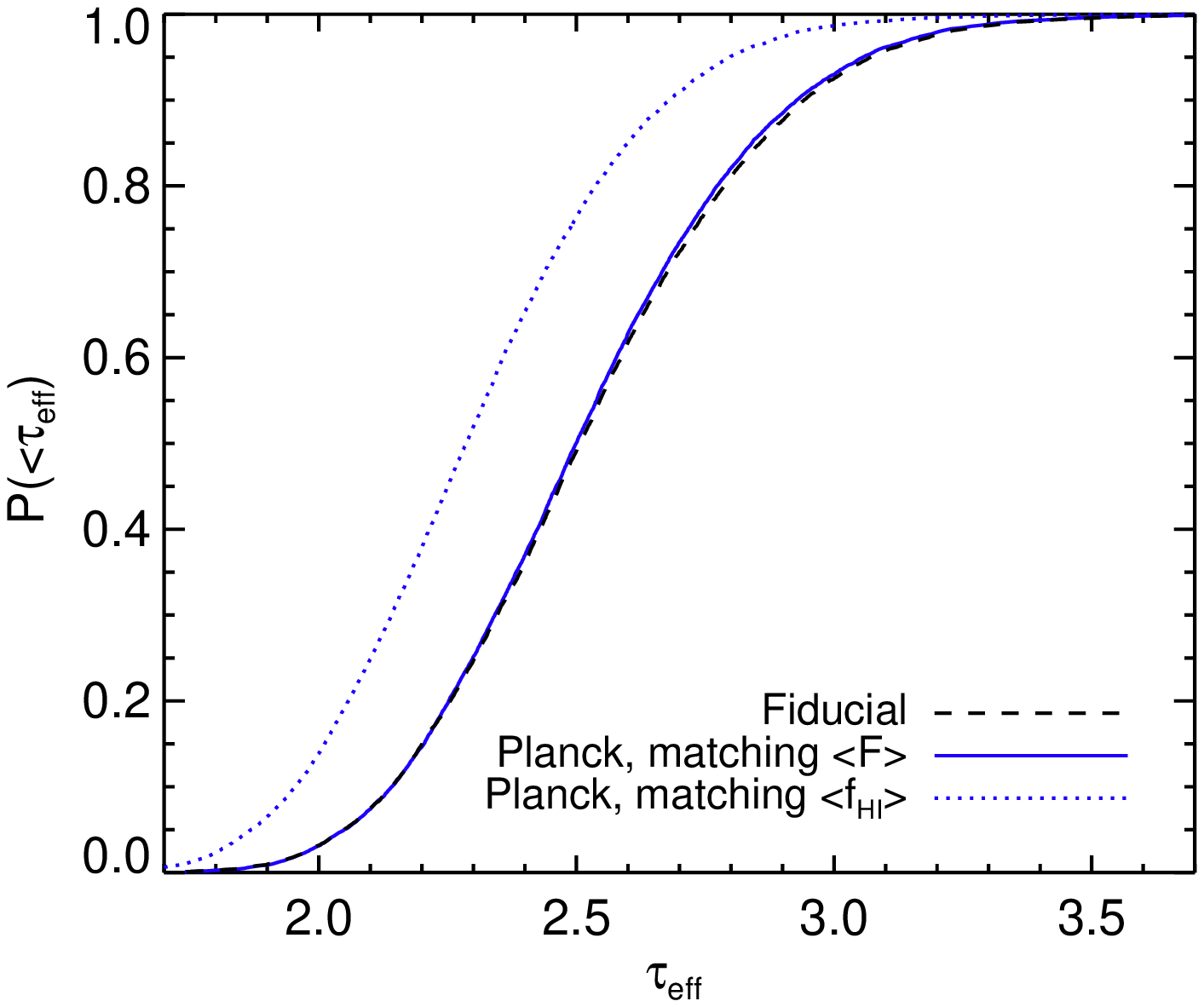}
   \vspace{-0.05in}
   \caption{Dependence of \Ptaueff\ on cosmology.  The dashed line shows \Ptaueff\ for 50~\hinvMpc lines of sight drawn from the 100--1024 run using our fiducial simulation cosmology.  The photoionization rate has been tuned such that $\langle F \rangle = 0.084$ and $\langle f_{\rm H\,I} \rangle = 2.4 \times 10^{-5}$.  The solid and dotted lines show \Ptaueff\ for the same box size and mass resolution but using {\it Planck} cosmology.  The solid line shows \Ptaueff\ with $\langle F \rangle$ matching the fiducial case, while the dotted line shows \Ptaueff\ when matching in \fhi. }
   \label{fig:taueff_cosmology}
   \end{center}
\end{figure}

Our fiducial simulations use a cosmology with $(\Omega_{\rm m}, \Omega_{\Lambda}, \Omega_{\rm b}h^2, h, \sigma_8, n_{\rm s}) = (0.26, 0.74, 0.023, 0.72, 0.80, 0.96)$.  To test our sensitivity to cosmological parameters we ran an additional 100~\hinvMpc, $2 \times 1024^3$ simulation using $(\Omega_{\rm m}, \Omega_{\Lambda}, \Omega_{\rm b}h^2, h, \sigma_8, n_{\rm s}) = (0.308, 0.692, 0.0222, 0.678, 0.829, 0.961)$, consistent with the recent results from {\it Planck} \citep{planckXVI}.  At fixed $\langle F \rangle$ we find a negligible difference in \Ptaueff\ (Fig.~\ref{fig:taueff_cosmology}); however, the {\it Planck} cosmology has a 16 per cent higher neutral fraction.  Our results for \fhi\ (Fig.~\ref{fig:fHI}) include this correction.

\subsection{Thermal history}\label{app:Thistory}

\begin{figure}
   \begin{center}
   \includegraphics[width=0.45\textwidth]{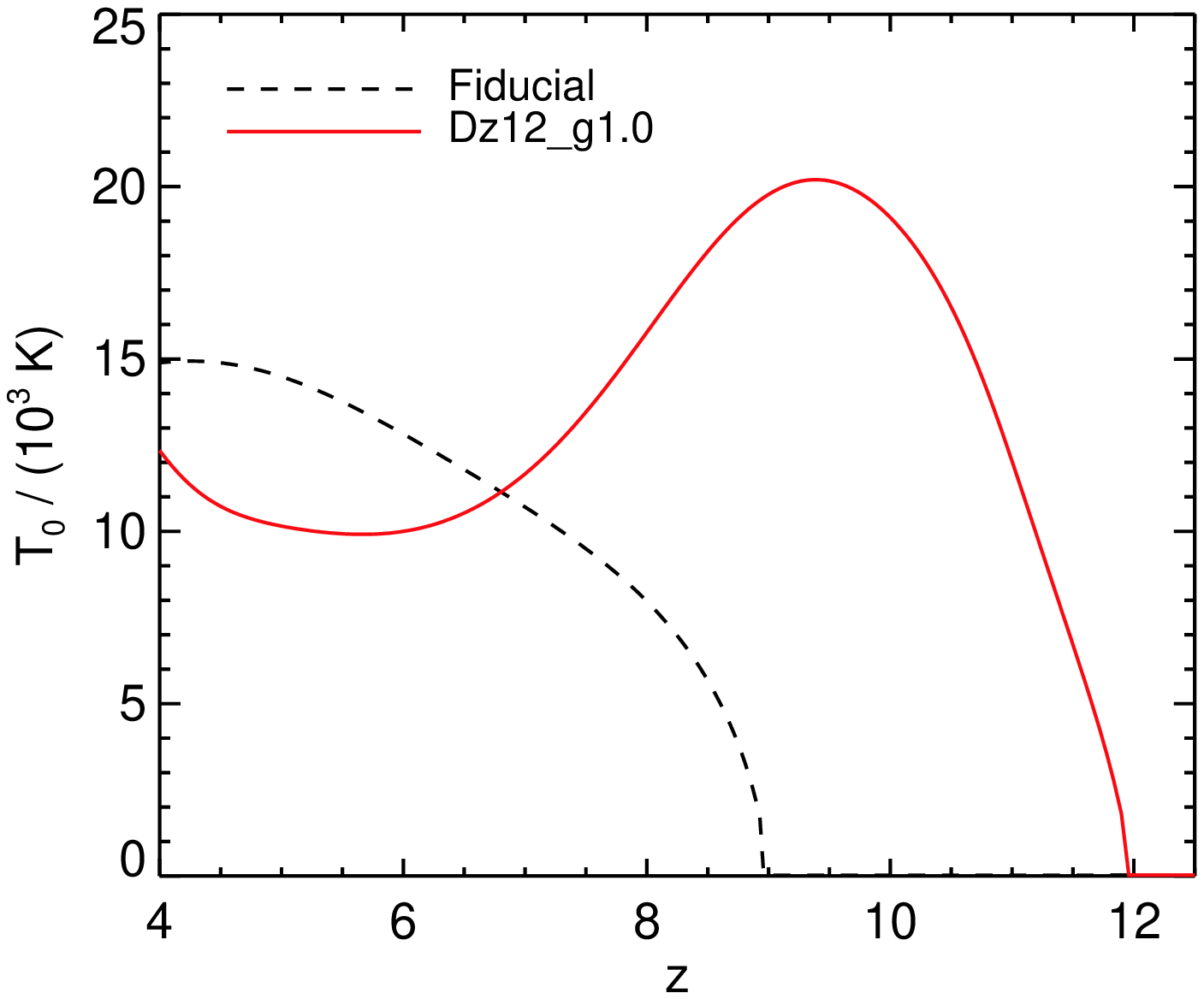}
   \vspace{-0.05in}
   \caption{Thermal histories for simulation runs used to probe the impact of Jeans smoothing on \Ptaueff\ for a uniform UVB.  The dashed line shows the temperature at the mean density as a function of redshift for our fiducial thermal history, while the solid line is for a case that invokes earlier photoionization heating and a nearly isothermal ($\gamma=1$) temperature-density relation.}
   \label{fig:Thist}
   \end{center}
\end{figure}

\begin{figure}
   \begin{center}
   \includegraphics[width=0.45\textwidth]{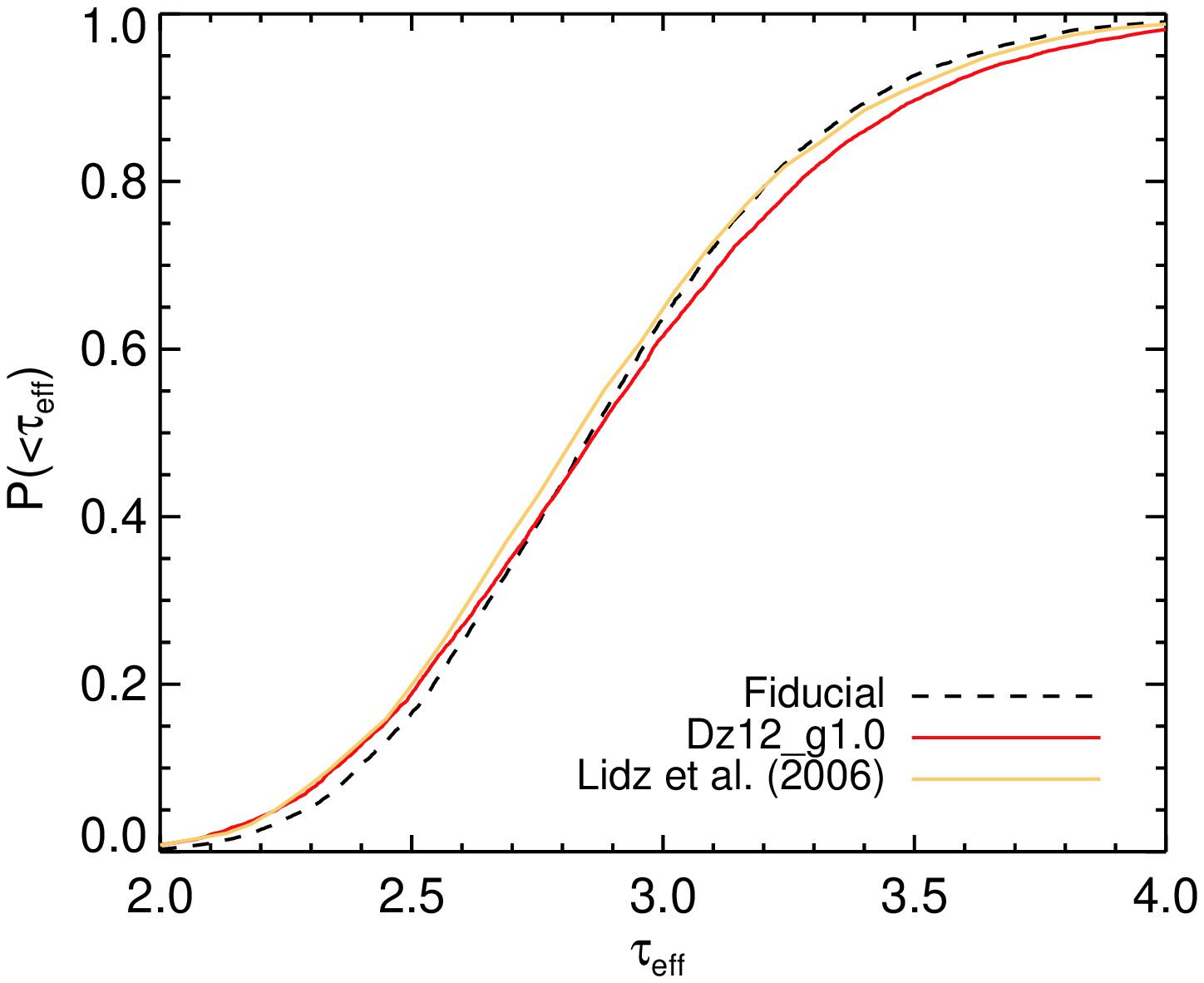}
   \vspace{-0.05in}
   \caption{Dependence of \Ptaueff\ on the thermal history of the IGM.  The dashed and red solid lines show \Ptaueff\ at $z=5.62$ for the corresponding thermal histories plotted in Fig.~\ref{fig:Thist}.  In this case we have computed \taueff\ over 40~\hinvMpc\ lines of sight and rescaled the photoionization rates to produce $\langle F \rangle = 0.06$ in order to compare with \Ptaueff\ from \citet{lidz2006}, shown as a yellow solid line.  We note that the Lidz et al. \Ptaueff\ is computed at $z=5.7$; however, we find that small differences in redshift have relatively little impact provided that $\langle F \rangle$ remains fixed.}
   \label{fig:taueff_Thist}
   \end{center}
\end{figure}

The thermal history of the IGM may also impact \Ptaueff.  Greater heating during hydrogen reionization, for example, can suppress the accretion of mass on to low-mass haloes, leaving more gas in the voids.  We tested this affect by running our 100--1024 simulation with two thermal histories, which are shown in Fig.~\ref{fig:Thist}.  In our fiducial run, the gas is reionized at $z_{r} = 9$ and allowed to heat up gradually.  In this run we use a temperature-density relation $T = T_{0} (\rho/\langle \rho \rangle)^{\gamma-1}$ with $\gamma \simeq 1.4$ at $z \sim 6$.  Run Dz12\_g1.0, in contrast, reionizes earlier ($z_{\rm r} = 12$), heats the gas more strongly at reionization, and uses $\gamma = 1.0$, which increases the heating in the voids.  We find a somewhat broader \Ptaueff\ in this run, although the difference is not large (Fig.~\ref{fig:taueff_Thist}).  For this test we compute \taueff\ over 40~\hinvMpc\ regions in order to facilitate a direct comparison with the results of \citet{lidz2006}.  The thermal history for run Dz12\_g1.0 is comparable to that used by Lidz et al., and we find a similar \taueff\ distribution.

\subsection{Galaxy UVB parameters}\label{app:uvb_convergence}

\begin{figure}
   \begin{center}
   \includegraphics[width=0.45\textwidth]{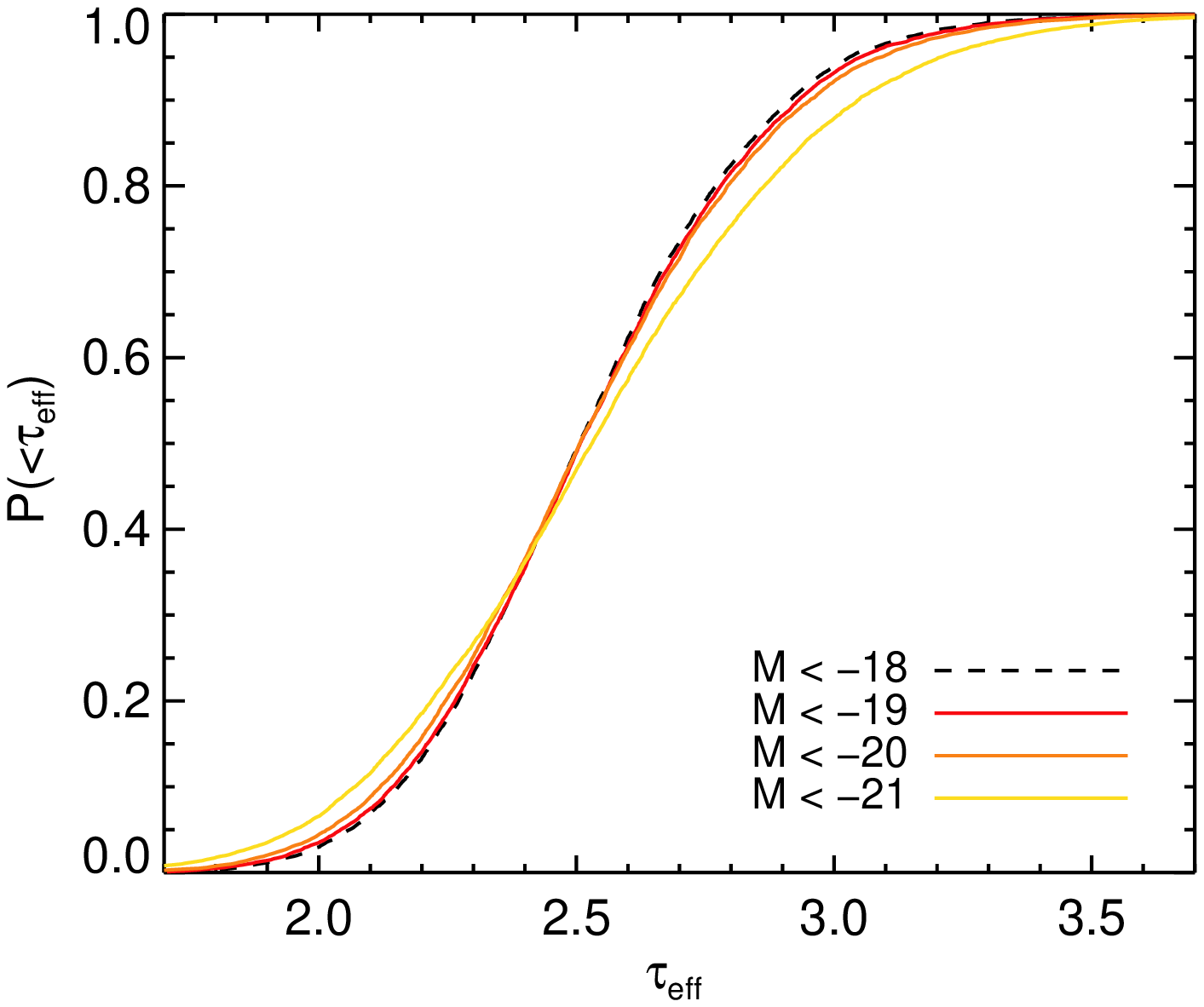}
   \vspace{-0.05in}
    \caption{Dependence of \Ptaueff\ in our galaxy UVB models on the limiting absolute magnitude of the sources.  In each case we compute \taueff\ values at $z=5.62$ over 50~\hinvMpc\ lines of sight after scaling the total ionizing emissivity such that $\langle F \rangle = 0.084$.  The width of \Ptaueff\ increases as the sources become increasingly rare.  At our fiducial cutoff ($M_{\rm AB} \le -18$), however, \Ptaueff\ appears to be well converged.}
   \label{fig:taueff_Mmax}
   \end{center}
\end{figure}

Out fiducial galaxy UVB models presented in Section~\ref{sec:gal_uvb} integrate over the ionizing emissivity from galaxies with $M_{\rm AB} \le -18$.  In principle this cutoff may cause us to overestimate the scatter in \taueff\ since we are neglecting contributions from fainter galaxies that are less biased with respect to the density field.  To estimate the magnitude of this effect we calculated our UVB at $z = 5.62$ while varying the upper limit in $M_{\rm AB}$ from $-$21 to $-$18, adjusting \fion\ to achieve the same mean transmitted \lya\ flux in each case.  The results for \Ptaueff\ are shown in Fig.~\ref{fig:taueff_Mmax}.  As expected, models that include only contributions from rarer, brighter galaxies, which we assign to more massive haloes, show a broader range in \taueff.  We find, however, that \Ptaueff\ is essentially converged when integrating up to $M_{\rm AB} = -19$.  Decreasing the galaxy duty cycle from unity essentially pushes the sources down to lower mass haloes, which we find has little affect on \Ptaueff.  We also find no dependence on the assumed galaxy UV spectral slope. 

\subsection{Ly$\alpha$/Ly$\beta$ ratio}\label{app:lyb_convergence}

\begin{figure}
   \begin{center}
   \includegraphics[width=0.45\textwidth]{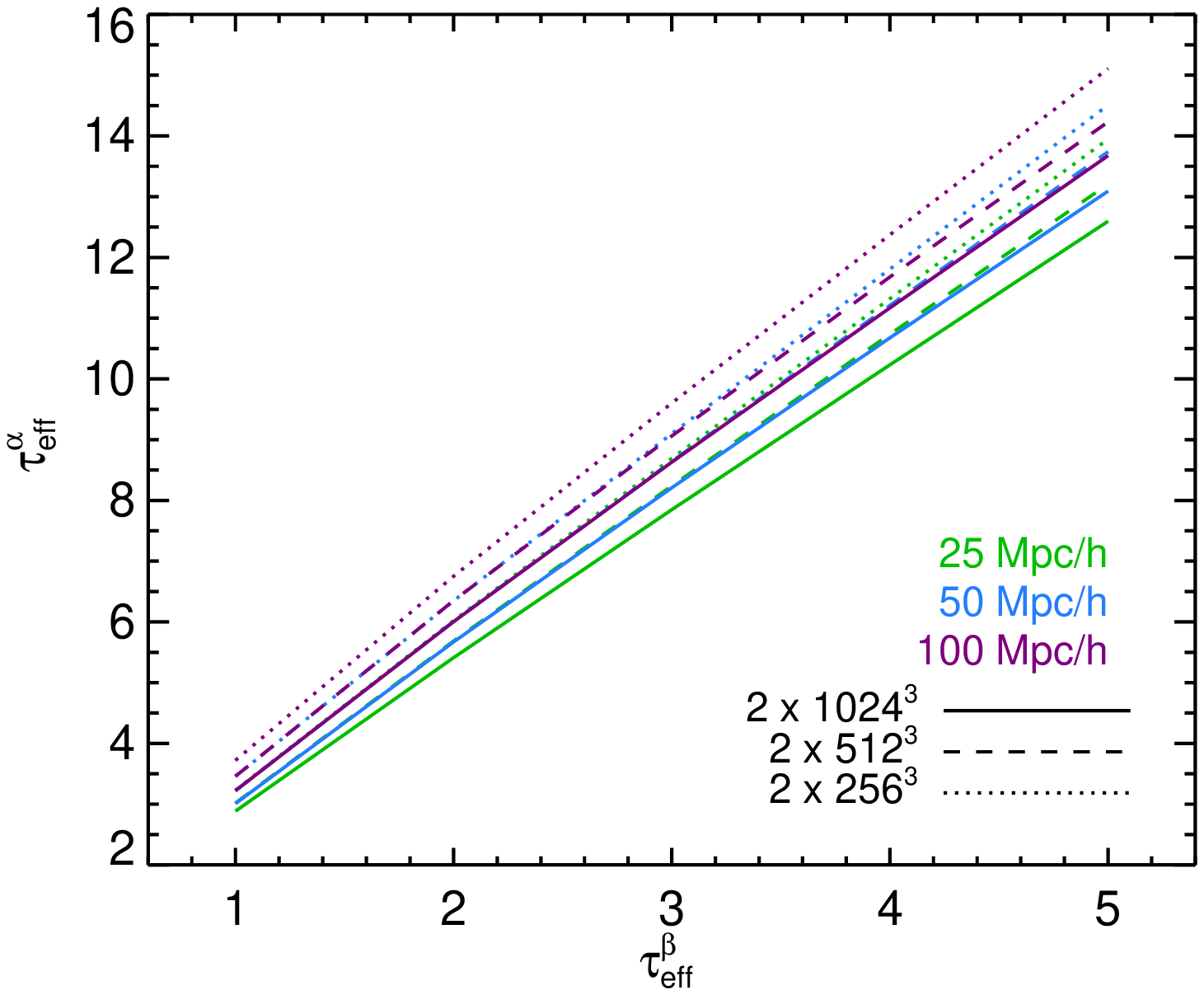}
   \vspace{-0.05in}
    \caption{Convergence of relationship between \taueffa\ and \taueffb\ with box size and resolution.  Colours and line styles denote different box sizes and particle numbers, as in Fig.~\ref{fig:neutralfrac_convergence}.  For each combination of these parameters we compute \taueffa, averaged over 5000 lines of sight at $z=5.62$, as a function of \taueffb.  Different values of \taueffb\ are achieved by adjusting the photoionization rate and do not include a contribution from foreground \lya\ absorption.  The relationship between \taueffa\ and \taueffb, which probe different density ranges, depends relatively little on box size at a fixed mass resolution; however, it is not well converged with mass resolution over the range covered here.}
   \label{fig:lyb_convergence}
   \end{center}
\end{figure}

Finally, we examine the dependence of the relationship between
\lya\ and \lyb\ opacity on box size and mass resolution.  Since
\lyb\ effectively probes higher density gas, \taueffb\ is expected to
converge more quickly than \taueffa\ in SPH simulations.  Moreover,
since \lya\ and \lyb\ probe different density ranges, the predicted
\taueffa\ at a fixed \taueffb\ may depend on the simulation
parameters. This effect is demonstrated in
Fig.~\ref{fig:lyb_convergence}, where we plot \taueffa\ as a
function of \taueffb\ for our nine convergence test runs.  Box size
has relatively little effect; however, \taueffa\ is lower in runs with
finer mass resolution.  This is again due to the fact that the centres
of voids are more highly evacuated, and therefore more transparent, in
runs that use a smaller gas particle mass.  This has a greater impact
on \lya\ than on \lyb.  We note that we have neglected foreground
\lya\ absorption in the \lyb\ forest for this test.  Our upper limit
for \taueffa\ for the trough in \ulas0148\ based on the \lyb\ opacity,
for which we used the 25--1024 run, should
nevertheless be conservative in terms of numerical convergence.

\end{document}